\documentclass[11pt,a4paper]{article}
\usepackage[utf8]{inputenc} 
\usepackage[english]{babel}
\usepackage[T1]{fontenc}
\usepackage{graphicx,amssymb,amsmath,wasysym}
\usepackage{hyperref} 
\usepackage{feynmf} 

\newcommand{\s}{\smallskip} 
\def\gtilu{{\tilde g}_u}
\def\gtild{{\tilde g}_d}
\def\gtilup{{\tilde g}_u^{\,\prime}}
\def\gtildp{{\tilde g}_d^{\,\prime}}
\def\mgut{M_{\rm \scriptscriptstyle GUT}}
\def\msusy{M_S}
\def\tb{\tan\beta}

\begin{document}
\begin{fmffile}{pepe}
\baselineskip=15.5pt

\thispagestyle{empty}

\begin{flushright}
UB-ECM-PF-09/14\\
ICCUB-09-205\\
\end{flushright}

\vspace{.5cm}

\begin{center}

{\Large\sc{\bf Dark matter direct detection in the MSSM with heavy scalars}}

\vspace*{9mm}
\setcounter{footnote}{0}
\setcounter{page}{0}
\renewcommand{\thefootnote}{\arabic{footnote}}

\mbox{ {\sc Nicolás BERNAL}}

\vspace*{0.9cm}

{\it High Energy Physics Group, Dept. ECM, and Institut de Ciències del Cosmos\\
Univ. de Barcelona, Av. Diagonal 647, E-08028 Barcelona, Catalonia, Spain}\s

\end{center}

\vspace{1cm}

\begin{abstract}
We explore the dark matter detection prospects in the Minimal Supersymmetric
Standard Model in the scenario where the scalar partners of the
fermions and the Higgs particles (except for the Standard-Model-like
one) are assumed to be very heavy and are removed from the low-energy
spectrum.
We analyse the neutralino LSP ($\chi_1^0$)
in scenarios where the gaugino mass parameters are universal at the GUT scale and also the case where they are non-universal.
This analysis is carried out in the framework of a Xenon-like $100$ kg experiment.
In general, an important fraction of the parameter space giving rise to the dark matter relic density measured by WMAP can be probed and excluded in the case of not detecting any WIMP.
In the opposite case, once a WIMP signal has been found, we show that for a light $\chi_1^0$ which is a higgsino-gaugino mixture it is possible to reconstruct efficiently the mass and the scattering cross-section of the neutralino LSP.
Moreover, we show that it is also feasible to put strong constraints over some of the parameters of the Lagrangian, e.g. the higgsino and the gaugino mass parameters.
\end{abstract}

\newpage

\section{Introduction}
There exists strong evidence that a large fraction of the Universe is dark and non-baryonic \cite{Jungman:1995df,Munoz:2003gx,Bertone:2004pz}.
Weakly Interacting Massive Particles (WIMPs), with masses lying from the GeV to the TeV scale, are the leading cold Dark Matter (DM) candidates.
The direct detection of WIMPs could not only directly confirm the existence of dark matter but would also probe the fundamental parameters of the underlying theory.
Constraints on, or measurements of, the WIMP mass and scattering cross-section will be complementary to the information derived from different experiments such as colliders \cite{Djouadi:2005gj,Baltz:2006fm,Carena:2006nv}, neutrino detectors \cite{Mena:2007ty}, dark matter indirect detection \cite{Dodelson:2007gd} or even other dark matter direct detection experiments \cite{Green:2007rb,Green:2008rd,Drees:2008bv,Drees:2009tr}.
It is therefore pertinent to examine the accuracy with which DM direct detection experiments will be able to measure the WIMP properties, if they are detected.\s

Low-energy supersymmetry (SUSY) with $R$-parity, and in particular the Minimal Supersymmetric Standard Model (MSSM), provides several well-motivated WIMP candidates which take the form of the Lightest Supersymmetric Particle (LSP) (see e.g. reference \cite{Jungman:1995df}).
However, no SUSY particle has been found at particle colliders and no clear indication of SUSY has emerged from any of the current measurements.
In this way, the initial theoretical motivations for SUSY, although still very appealing, are being questioned and the scale of new physics beyond the Standard Model (SM) is slowly, but steadily, drifting above the weak scale.
In this vein, a SUSY framework named \textit{Split Supersymmetry} \cite{ArkaniHamed:2004fb,Giudice:2004tc,ArkaniHamed:2004yi} has emerged, proposing a generic realisation of SUSY, where the scalar superpartners of SM fermions are extremely heavy, with a mass that in principle can reach the GUT scale.
Moreover, the fermionic superpartners of the Higgs and gauge bosons could remain near the electroweak scale, protected by symmetries.
Let us note that this type of models may appear naturally in some string inspired models with F-term uplifting \cite{Dudas:2006gr,Lebedev:2006qq,Lebedev:2006qc}.
In Split SUSY, two major features of the MSSM are maintained: on the one hand the unification of gauge couplings works essentially in the same way as in the MSSM.
Actually, it has been pointed out \cite{ArkaniHamed:2004fb,Giudice:2004tc} that, in order for SUSY to provide solutions to the unification problem, only gauginos and higgsinos, the fermionic superpartners of the gauge and Higgs bosons, need to be relatively light.
On the other hand, in Split SUSY the lightest neutralino is usually the LSP, with a mass lying from few dozen GeV to the TeV scale, hence being a promising candidate for the cold dark matter.
It is clear that in this limit where scalars are ultra heavy, the fine-tuning problem is reintroduced in the theory.\s

However, it is not mandatory to have scalar superpartners at such a high scale.
Indeed, from a scalar mass $\msusy$ of the order of $\sim 10^4$ GeV the scalar particles are decoupled from the low-energy spectrum, and hence their interference with low-energy phenomenology is marginal.
In this case, and as compared to the MSSM, just a slightly large amount of fine-tuning is needed.
Let us recall that there is no compelling criterion to define the maximal acceptable amount of fine tuning \cite{Barbieri:1987fn}, and the choice of the upper bound on the SUSY scale $\msusy$ is somewhat subjective.
An interesting feature of such a MSSM with heavy scalars \cite{Bernal:2007uv}, is that it is much more predictive than the general MSSM, since a large number of the free parameters comes from the scalar sector.
At the same time, both the gauge couplings unification and the solution to the dark matter problem are maintained.\s

The present article is devoted, on the one hand, to the analysis of the neutralino--nuclei scattering cross-section and to the study of prospects of neutralino dark matter direct detection within the MSSM with heavy scalars. 
Some previous studies \cite{Pierce:2004mk,Masiero:2004ft,Beylin:2009wz} have already been devoted to these subjects considering low-energy model-independent frameworks.
Instead of following the aforementioned approach, we will study the general model using different patterns of soft SUSY-breaking gaugino masses.
In this way, we will consider different SUSY models in which the boundary conditions for the gaugino masses at the GUT scale are different from those of the universal scenario.\\
On the other hand, we will study the possibilities offered by direct detection experiments to reconstruct: i) the mass and the scattering cross-section \cite{Green:2007rb,Green:2008rd,Bernal:2008zk} and ii) the fundamental parameters of the Lagrangian.
The present paper aims to provide a complementary picture to that use in references \cite{:2008gva,Turlay:2008dm} and \cite{Bernalilc}, in which the reconstruction of the MSSM with heavy scalars is studied in the framework of high-energy colliders, both at LHC and at ILC.

The rest of the paper is organized as follows.
In the next section we briefly review the MSSM with heavy scalars and discuss the neutralino dark matter within this model.
In section \ref{dd}, we review the spin-independent direct detection techniques, focusing on the event rate and the WIMP-nucleon
scattering cross-section for a Xenon-like experiment, in a microscopically model independent approach.
This analysis lays the basis for the following study, in the framework of the MSSM with heavy scalars.
In section \ref{ddp} we present some results of dark matter direct detection prospects for a Xenon-like experiment.
Section \ref{rp} is dedicated to the possibilities of reconstructing both physical observables and fundamental parameters of the model.
Before closing, conclusions are given in section \ref{conclu}.

\section{The MSSM with heavy scalars}
\subsection{Definition of the model}
In the MSSM with heavy scalars, the scalar superpartners of the leptons and quarks as well as most of the Higgs bosons are assumed to be very heavy, at a common mass scale $\msusy\gg 1$ TeV.
The low-energy theory contains, besides the SM-like Higgs boson only the two higgsinos ($\tilde H_u$, $\tilde H_d$) and the three gauginos: the bino ($\tilde B$), the wino ($\tilde W$) and the gluino ($\tilde g$). 
Omitting the gauge-invariant kinetic terms, as well as non-renormalizable
operators suppressed by powers of the heavy scale $\msusy$, the Lagrangian of the effective theory
reads \cite{ArkaniHamed:2004fb,Giudice:2004tc}
\begin{eqnarray} \label{lagrangien}
{\cal L}&\supset&m_H^2 H^\dagger H-\frac{\lambda}{2}\left( H^\dagger H\right)^2
-\biggr[ h^u_{ij} {\bar q}_j u_i\epsilon H^*
+h^d_{ij} {\bar q}_j d_iH
+h^e_{ij} {\bar \ell}_j e_iH \nonumber \\
&+&\frac{M_3}{2} {\tilde g}^A {\tilde g}^A
+\frac{M_2}{2} {\tilde W}^a {\tilde W}^a
+\frac{M_1}{2} {\tilde B} {\tilde B}
+\mu\, {\tilde H}_u^T\epsilon\, {\tilde H}_d\nonumber\\
&+&\frac{H^\dagger}{\sqrt{2}}\left( \gtilu \sigma^a {\tilde W}^a
+\gtilup {\tilde B} \right) {\tilde H}_u
+\frac{H^T\epsilon}{\sqrt{2}}\left(
-\gtild \sigma^a {\tilde W}^a
+\gtildp {\tilde B} \right)
{\tilde H}_d +{\rm h.c.}\biggl]\,,
\end{eqnarray}
where $\sigma^a$ are the Pauli matrices, $\epsilon\equiv i\,\sigma^2$ and $i$, $j$ are generation indices.
The SM-like Higgs
doublet $H$ is a linear combination of the two MSSM Higgs doublets $H_u$ and $H_d$, which needs to be properly fine-tuned
to have a small mass term $m_H^2$:
\begin{equation}
H=-\cos\beta\,\epsilon\,H_d^*+\sin\beta\,H_u\,.
\end{equation}
At the high scale $\msusy$ the boundary conditions on the quartic Higgs coupling and on the
Higgs-higgsino–gaugino couplings of the effective theory are determined by SUSY invariance and yield:
\begin{equation}
\lambda(\msusy)=\frac14\left[g^2(\msusy)+g'^2(\msusy)\right]\,\cos^22\beta,
\end{equation}
\begin{eqnarray}
\label{boundlam}
\gtilu(\msusy ) = g (\msusy )\sin\beta&,& \hspace*{1cm}
\gtild(\msusy ) = g (\msusy )\cos\beta~, \\
\gtilup(\msusy )= g^{\,\prime} (\msusy ) \sin\beta&,& \hspace*{1cm}
\gtildp(\msusy )= g^{\,\prime} (\msusy )\cos\beta~,
\end{eqnarray}
where $g$ and $g^{\,\prime}$ are the $SU(2)_L$ and $U(1)_Y$ gauge couplings.
$\tb$ is interpreted as the angle that rotates the two Higgs doublets into one heavy and one light (SM-like).\s


Since for the MSSM with heavy scalars the number of basic input parameters is rather small, one can relax the assumption of a universal gaugino mass at $\mgut$ and still have a predictive model which could lead to a different phenomenology with respect to the universal scenario.
In reference \cite{Bernal:2007uv}, various scenarios for non-universal gaugino masses have been discussed.
A particular interesting one is the gravity-mediated SUSY-breaking scenario in
which the gaugino masses arise from a dimension-$5$ operator
${\cal L}
\propto \langle F_\Phi \rangle _{ab} / M_{\rm Pl}  \cdot \lambda^a \lambda^b $,
where  $\lambda$ are the gaugino fields and $F_\Phi$ is the auxiliary component of a left-handed chiral superfield
$\Phi$ which couples to the SUSY field strength. In the usual minimal Supergravity (mSUGRA) model with $SU(5)$ grand unification,  the
field $F_\Phi$ is a singlet under the unifying gauge group, leading to universal
gaugino masses. However, the field $\Phi$ could sit in any representation of the
symmetric product of the adjoint group \cite{Ellis:1985jn}. For $SU(5)$ symmetry,
$F_\Phi$ could belong to an irreducible representation which  results from the
symmetric product of two adjoints
\begin{equation}
(\mathbf{24 \otimes 24})_{\text symmetric}=\mathbf{1\oplus 24\oplus 75\oplus 200}\,.
\end{equation}
Once the neutral component of $F_\Phi$ has acquired a vev, $\langle F_\Phi
\rangle _{ab}= V_{a} \delta_{ab}$, the vevs $V_a$ determine the relative
magnitude of the soft SUSY-breaking gaugino mass parameters $M_a$ at $M_{\rm
GUT}$ \cite{Amundson:1996nw,Anderson:1996bg}.
This is shown in the left-hand side of table \ref{tab-gauginos}
and, as can be seen, only in the singlet case {\bf 1} these parameters are universal.
\begin{table}[ht!]
\begin{center}
\renewcommand{\arraystretch}{1.3}
\begin{tabular}{|c||c|c|} \hline
$\ \ \ ~ \ \ \ $ & $\ \ Q=\mgut \ \ $ &
$\ \ Q=M_Z \ \ $ \\ \hline\hline
{\bf 1}  & $1 \,:\, 1\,:\, 1$ & $1.0\,:\, 2.0\,:\,  6.2$\\ \hline
{\bf 24} & $1 \,:\, 3\,:\,-2$ & $1.0\,:\, 6.0\,:\,-12.7$ \\ \hline
{\bf 75} & $5 \,:\,-3\,:\,-1$ & $1.0\,:\,-1.2\,:\, -1.3$ \\ \hline
{\bf 200}& $10\,:\, 2\,:\, 1$ & $2.5\,:\, 1.0\,:\,  1.6$ \\ \hline
\end{tabular}
\end{center}
\caption{The ratios of gaugino mass parameters, $M_1\!:\!M_2\!:\!M_3$,
at the renormalization scales $\mgut$ and $M_Z$,
for the different
patterns of soft SUSY breaking.
The values $\msusy= 10^4$ GeV and $\tb=5$ has been used.}
\label{tab-gauginos}
\end{table}
We also give the $M_a$ values at the
electroweak scale $M_Z$, by using the one-loop renormalization group equations.
In this vein, for a given representation besides the SM basic input parameters, the free parameters of the model are the common scalar mass $\msusy$, $\tb$, $\mu(M_Z)$ and $M_1$; $M_2$ and $M_3$ being fixed at $\mgut$.\s

For the determination
of the mass spectrum, the resummation
to all orders of the leading logarithms of the large scalar mass and the one-loop radiative corrections has been taken into account, using the 
routine {\tt SHeavy} of the code SuSpect \cite{Djouadi:2002ze} described in reference \cite{Bernal:2007uv}.

\subsection{Dark matter within the MSSM with heavy scalars}
As deduced from the WMAP satellite measurement of the temperature anisotropies in the 
Cosmic Microwave Background, cold dark matter makes up approximately $23\%$ of the energy
of the Universe \cite{Dunkley:2008ie}. The DM cosmological density is precisely measured to be
\begin{equation}
\Omega_{DM}\,h^2=0.1099\pm 0.062\,,
\end{equation}
which leads to $0.104\lesssim\Omega_{DM}\,h^2\lesssim 0.116$ at $68\%$ CL.
The accuracy is expected to be
improved to the percent level by future measurements at Planck satellite \cite{Bouchet:2007zz}.\s

As it is well known, the LSP neutralino can be an ideal WIMP-like cold DM candidate
and in some areas of the SUSY parameter space
the cosmological relic density associated to the lightest neutralino $\chi_1^0$, which is inversely
proportional to the neutralino annihilation cross-section $\sigma_{\rm
ann} \equiv \, \sigma(\chi_1^0 \chi_1^0 \rightarrow \text{SM\,
particles})$, falls into the range required by WMAP.
In the MSSM with heavy scalars, the neutralino mass matrix in the bino, wino, higgsino basis ($\tilde B$, $\tilde W_3$, $\tilde H^0_d$, $\tilde H^0_u$) reads as follows:
\begin{equation}
{\cal M}_N = \left( \begin{array}{cccc}
M_1 & 0 & -\frac{{\tilde g}'_d\,v}{\sqrt 2}
& \frac{{\tilde g}'_u\,v}{\sqrt 2} \\
0   & M_2 & \frac{{\tilde g}_d\,v}{\sqrt 2}
& -\frac{{\tilde g}_u\,v}{\sqrt 2} \\
-\frac{{\tilde g}'_d\,v}{\sqrt 2}
&\frac{{\tilde g}_d\,v}{\sqrt 2} & 0 & -\mu \\
\frac{{\tilde g}'_u\,v}{\sqrt 2} &
-\frac{{\tilde g}_u\,v}{\sqrt 2} & -\mu & 0
\end{array} \right)~,
\label{chimass}
\end{equation}
where $v\sim 174$ GeV is the SM-like Higgs boson vev.
Thus the lightest neutralino $\chi_1^0$ is a linear superposition of the bino, wino and higgsinos states:
\begin{equation}
\chi_1^0=N_{11}\,\tilde B+N_{12}\,\tilde W_3+N_{13}\,\tilde H^0_d+N_{14}\,\tilde H^0_u\,;
\end{equation}
the matrix $N$ rotates the neutralino states so that the
mass matrix $N\,{\cal M}_N\,N^T$ is diagonal.\s

In the MSSM with
heavy scalars, there are essentially only four regions in which the WMAP constraint is fulfilled:
\begin{itemize}

\item[--] The `mixed region' in which the LSP is a higgsino--gaugino
mixture \cite{ArkaniHamed:2006mb}, $M_1 \sim |\mu|$, which enhances (but not too much) its
annihilation cross-sections into final states containing gauge and/or
Higgs bosons and top quarks, $\chi_1^0 \chi_1^0 \to W^+ W^-$, $ZZ$, $HZ$,
$HH$ and $t\bar t$.

\item[--] The `pure higgsino' and `pure wino' regions, in which the
LSP is almost degenerate in mass with the lightest
chargino ($\chi_1^\pm$) and the next-to-lightest neutralino ($\chi_2^0$).
Such a scenario leads to an enhanced
annihilation of sparticles since the $\chi_1^0-\chi_1^\pm$ and $\chi_1^0-\chi_2^0$
coannihilation cross-sections \cite{Griest:1990kh} are much larger than that of the LSP.
This solution generally requires LSP masses beyond 1 TeV.

\item[--] The `$H$-pole' region in which the LSP is rather light,
$m_{\chi_1^0}\sim\frac12 M_H$, and the $s$-channel $H$ exchange is
nearly resonant allowing the neutralinos to annihilate efficiently \cite{Baer:2003ru,Djouadi:2005dz}.

\item[--] The `$Z$-pole' region in which the LSP is very light,
$m_{\chi_1^0}\sim\frac12 M_Z\sim 45$ GeV, and the $s$-channel $Z$ exchange is
nearly resonant.
We note that this region is not ruled out only in scenarios where the mass splitting between $M_1$ and $M_2$  at the electroweak scale is very large \cite{Bernal:2007uv}.

\end{itemize}

\begin{figure}[ht!]
\centering
\includegraphics[width=7.0cm,clip=true,angle=-90]{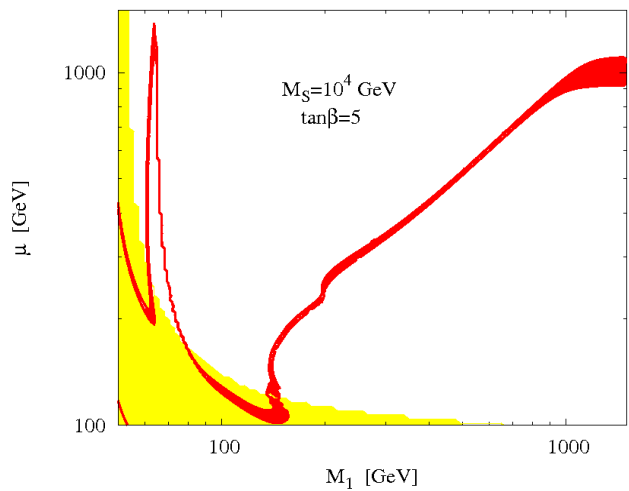}\vspace{-0.4cm}
\caption{The regions of the $[M_1,\,\mu]$ plane in which the WMAP constraint is
fulfilled (red--dark gray area) for a common scalar mass value $\msusy=10^4$ GeV, $\tb=5$ and the universal scenario. The green (light gray) area on the left and
the bottom is the one excluded by direct searches of SUSY particles.}
\label{dm4}
\end{figure}

The computation of dark matter relic density has been performed using an adapted version of the micrOMEGAs public code \cite{Belanger:2006is,Belanger:2008sj}, where all the scalar particles (except for the SM-like Higgs boson) have been integrated out.
Figure \ref{dm4} displays the area, in the $[M_1,\,\mu]$ plane,
in which the WMAP constraint is satisfied (red--dark gray);
a common scalar mass $\msusy=10^4$ GeV, $\tb=5$ and the universal scenario has been chosen.
Hereafter $M_1$ and $\mu$ have to
be interpreted as parameters evaluated at the electroweak scale.
The green (light grey) area in the
left and bottom parts of the figure denotes the region excluded by the
collider data.
This region is ruled out by the negative search of charginos at LEP2 from pair production of the lightest chargino: $e^+e^-\to\chi_1^\pm\chi_1^\mp$ \cite{Amsler:2008zzb}.
The peak for
small $M_1$ values at $M_1 \sim\frac12 M_H$, is due to the
$s$-channel exchange of an almost real Higgs boson, $\chi_1^0\,\chi_1^0 \to
H$. For the mass value obtained here, $M_H \sim 130$ GeV, the Higgs
boson mainly decays into $b\bar b$ final states.
Right into the peak
one is too close to the Higgs mass pole, and the LSP annihilation is
too efficient leading to a too small $\Omega_{DM}\,h^2$. The peak reaches up
to $\mu \sim 1300$ GeV, a value beyond which the LSP is almost
bino-like and its coupling to the Higgs boson is too small (the Higgs
prefers to couple to a higgsino--gaugino mixture) to generate a
sizable annihilation cross-section.\s

For larger $\mu$ and $M_1$ values there is an almost straight band in
which $\mu \sim M_1 \sim\frac12 M_2$ and the LSP is a bino--higgsino mixture with
sizable couplings to $W$, $Z$ and Higgs bosons, allowing for reasonably
large rates for neutralino annihilation into $\chi_1^0 \chi_1^0 \to
W^+W^-$, $ZZ$, $HZ$ and $HH$ final states. For instance, for $M_1\sim 200$
GeV and $\mu\sim 200$ GeV, the annihilation cross-section is mostly due to
the $WW$ and $HH$ final states and, to
a lesser extent, the $ZZ$ and $ZH$ final states. For slightly larger $\mu$ and $M_1$ values there is a jump due
to the opening of the $\chi_1^0 \chi_1^0 \to t \bar t$ channel, which
henceforth dominates the annihilation cross-section. Above and below the band,
the LSP couplings to the various final states are
either too strong or too weak to generate the relevant relic
density. For $\mu$ values close to $1$ TeV and even larger values of
$M_1$ there is a wider area in which the WMAP constraint is also
fulfilled. In this region the LSP is almost a pure higgsino and a
correct $\Omega_{DM}\,h^2$ can also be obtained thanks to the
coannihilation of the LSP with the $\chi_1^\pm$ and $\chi_2^0$
states.  For lower $\mu$ values and $M_1$ still very large the LSP
coannihilation with $\chi_1^\pm$ and $\chi_2^0$ is exceedingly strong and
leads to a too small $\Omega_{DM}\,h^2$.\s 

\begin{figure}[ht!]
\centering
\hspace{-0.7cm}\includegraphics[width=7.0cm,clip=true,angle=-90]{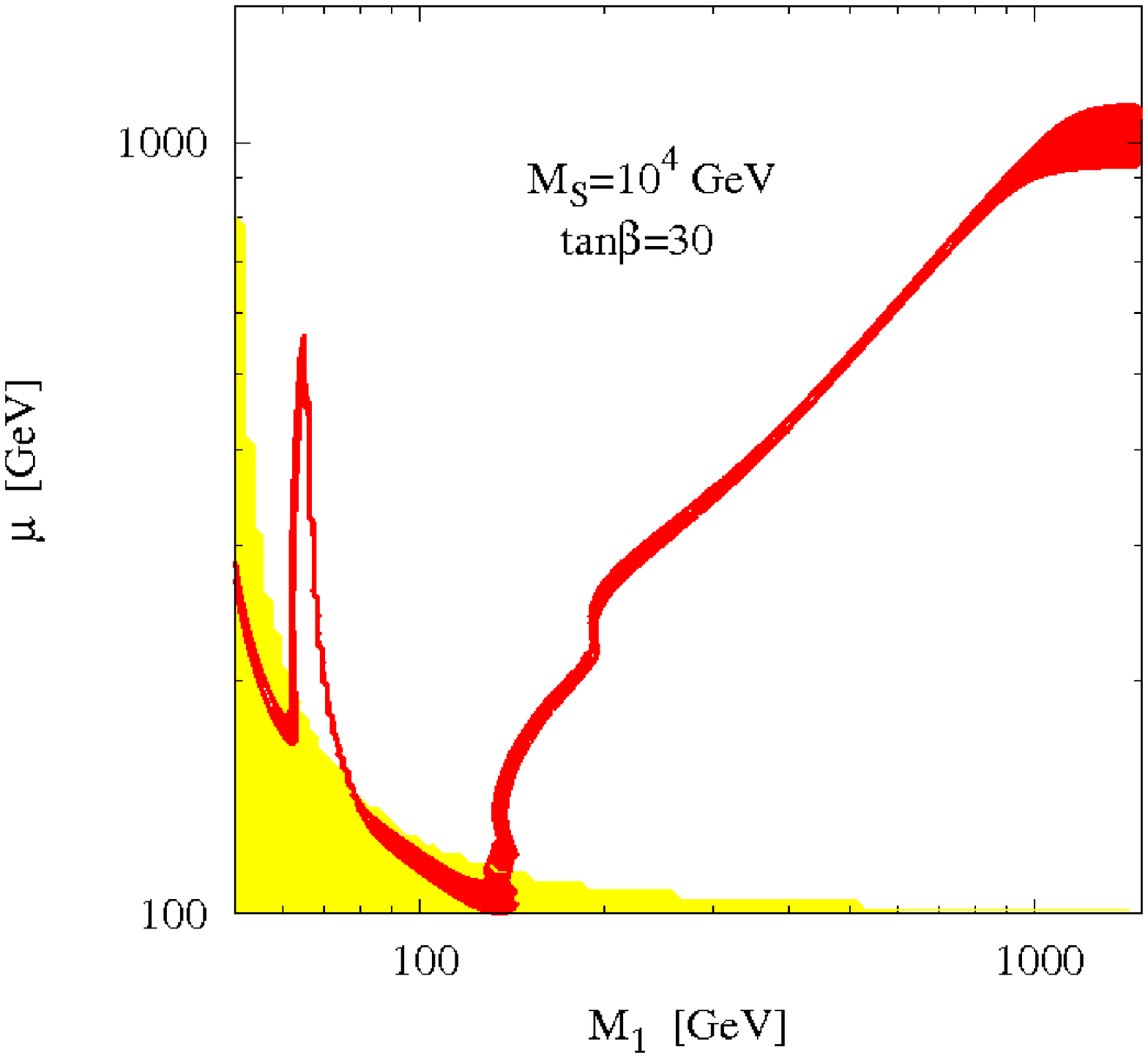}
\hspace{ 1.0cm}\includegraphics[width=7.0cm,clip=true,angle=-90]{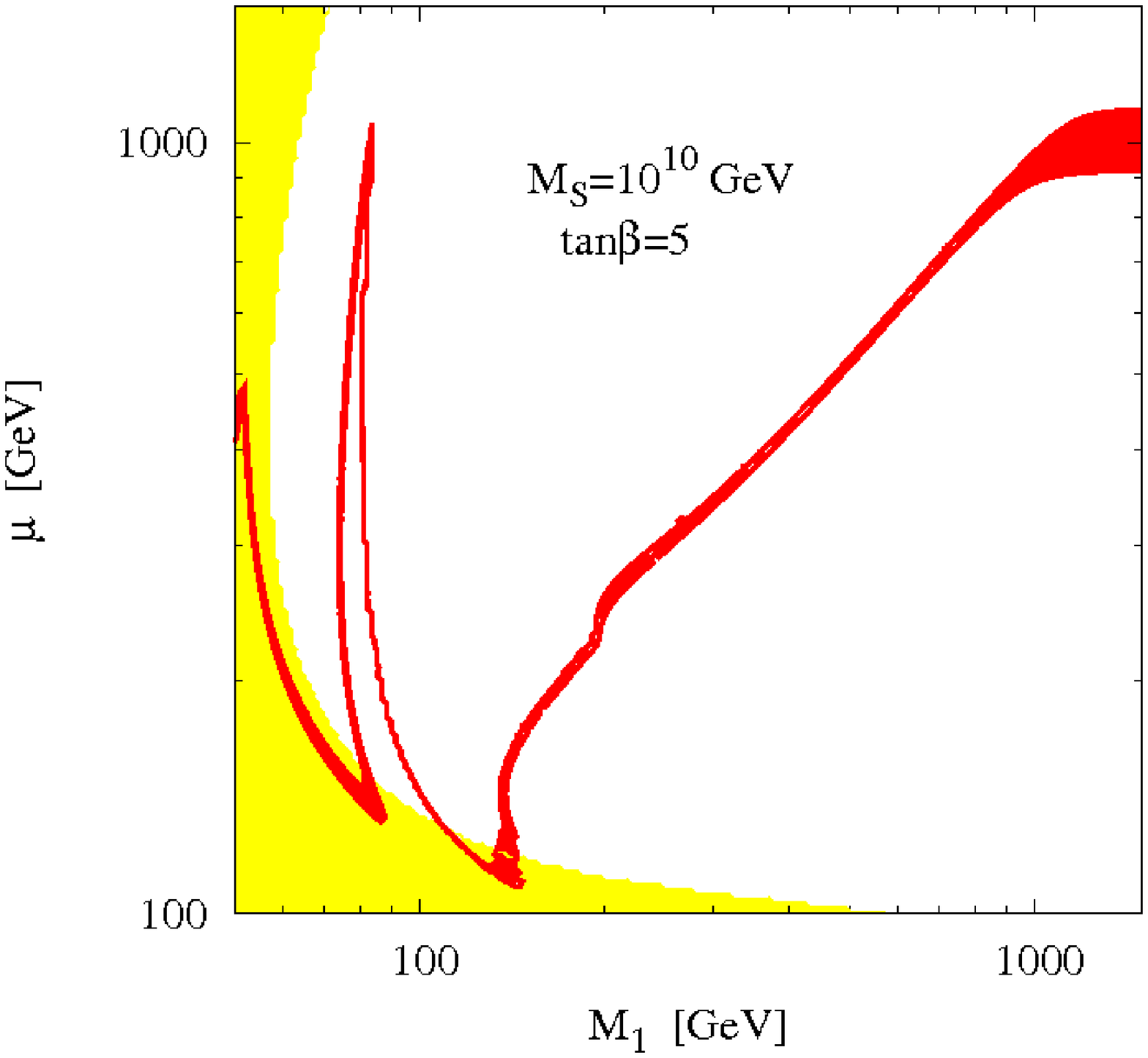}
\vspace{-0.4cm}
\caption{The same as in figure \ref{dm4} but for $\tb=30$ (left pane) or $\msusy=10^{10}$ GeV (right pane).}
\label{dmmas}
\end{figure}

In figure \ref{dmmas} we present a similar analysis to the previous one, with the only proviso that $\tb=30$ (left pane) and $\msusy=10^{10}$ GeV (right pane). 
The figure in the left-hand
side shows similar features to figure \ref{dm4}; a worth noting difference is that the peak which is due to the $s$-channel Higgs boson exchange reaches up to $\mu\sim 600$ GeV, owing to the fact that for small $\tb$ values the LSP
becomes bino-like faster than in the high $\tb$ case and its couplings to the Higgs boson
are thus smaller.
On the other hand, for $\msusy=10^{10}$ GeV, the Higgs peak is shifted to a slightly higher $M_1$ value, $M_1\sim\frac12 M_H\sim 80$ GeV.
In this
case the annihilation channel $\chi_1^0\chi_1^0\to H\to W\,W^*\to W\,f\bar f$
gives a significant contribution to the total cross-section \cite{Bernal:2007uv}.
Note that variations over $\msusy$ and $\tb$ are primarily reflected in the Higgs peak, whereas the mixed region is almost insensitive.\s

\begin{figure}[ht!]
\centering
\hspace{-0.7cm}\includegraphics[width=7.0cm,clip=true,angle=-90]{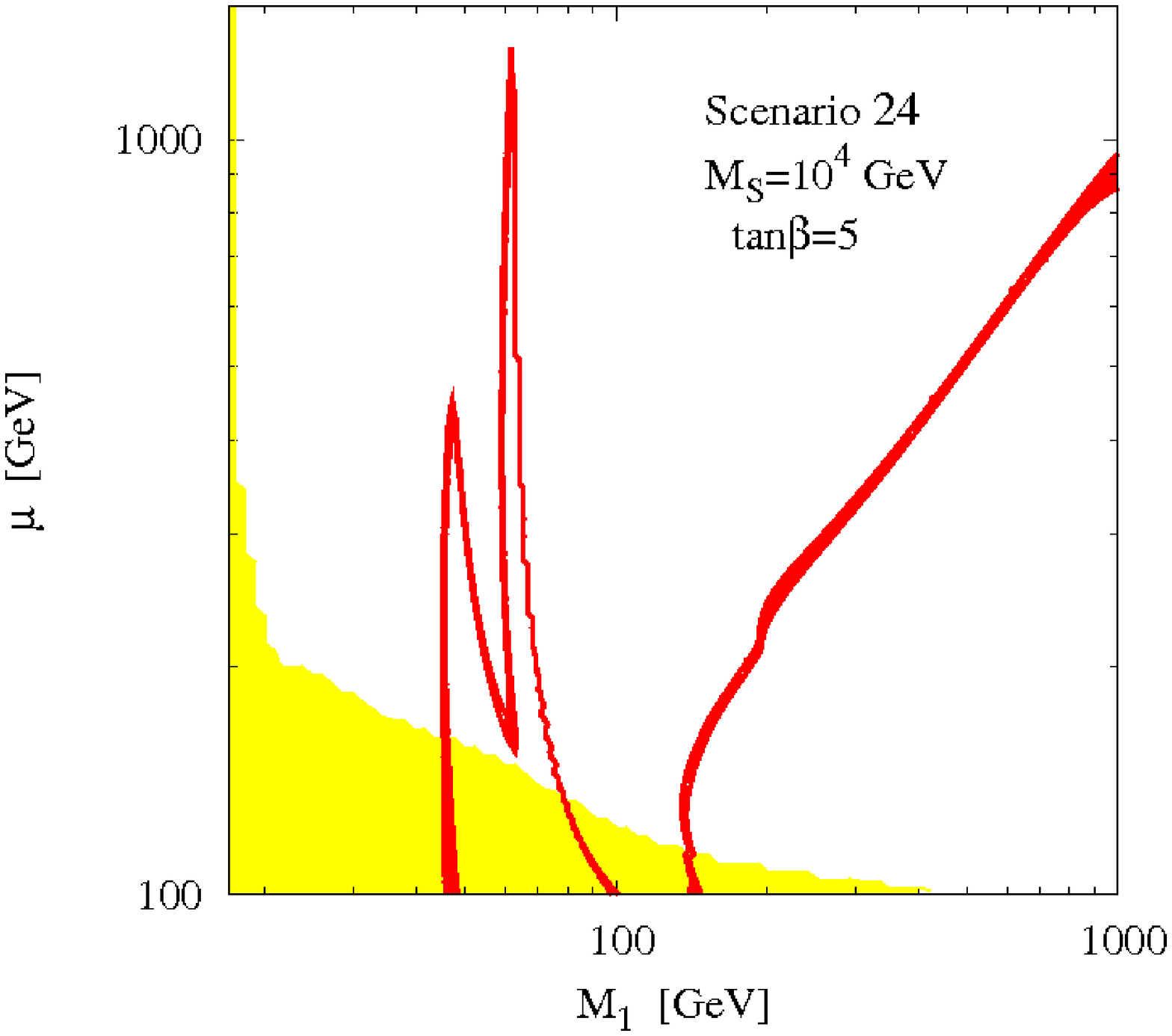}
\hspace{ 1.0cm}\includegraphics[width=7.0cm,clip=true,angle=-90]{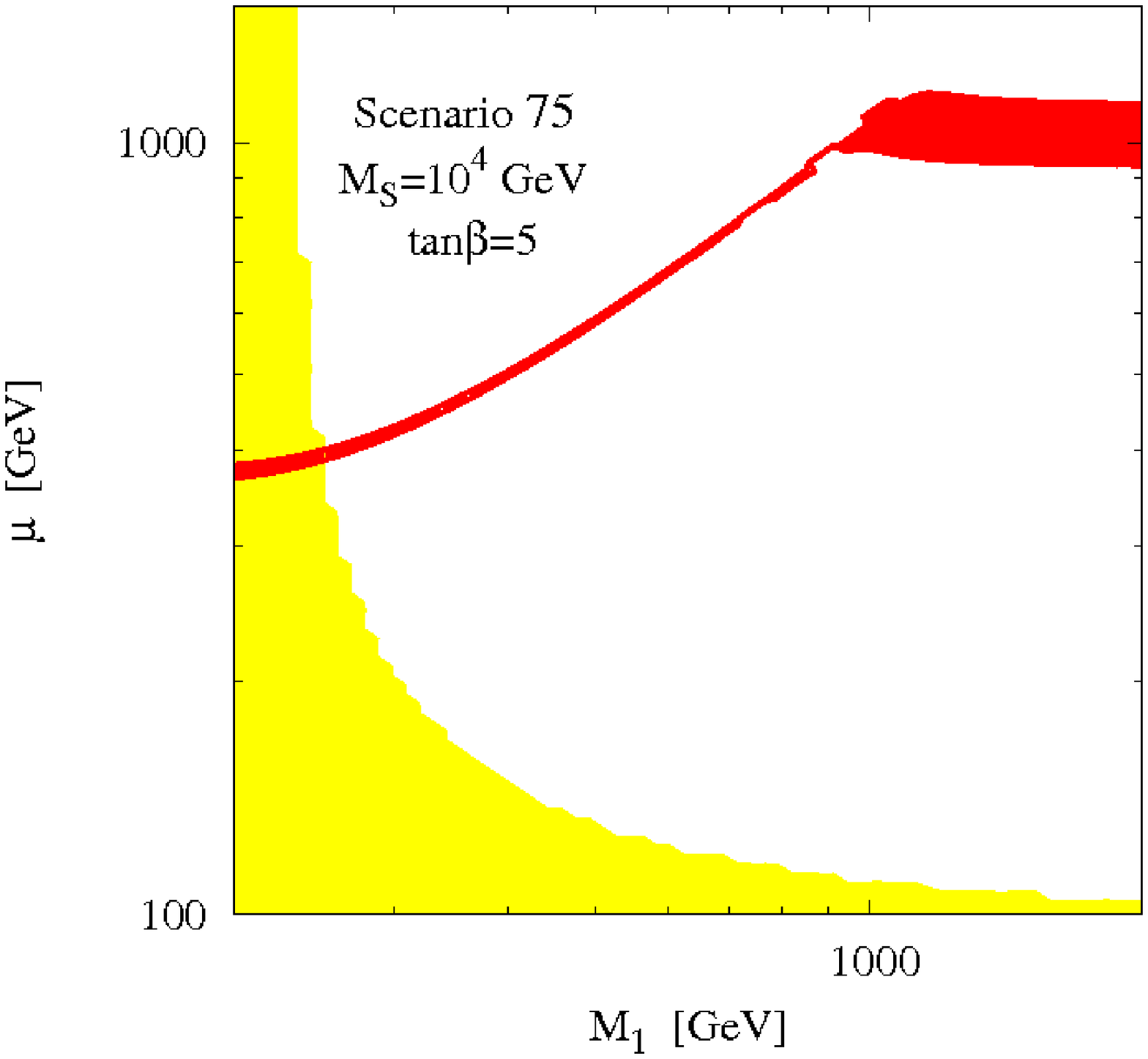}
\vspace{-0.4cm}
\caption{The same as in figure \ref{dm4} but for scenarios  {\bf 24} (left pane) and {\bf 75} (right pane).}
\label{dmnon}
\end{figure}

Figure \ref{dmnon} is similar to figure \ref{dm4}, with the difference that we take two situations presenting non-universality in the gaugino masses at $\mgut$: scenario {\bf 24} (left pane) and scenario  {\bf 75} (right pane).
The case considered in the left-hand
side shows similar features to the one with the universal setup explored in the foregoing, except that a new peak at $M_1\sim\frac12 M_Z\sim 45$ GeV appears, as a consequence of the $s$-channel $Z$ boson exchange.
In this scenario, such a resonance is no longer excluded because of the large gap between $M_1$ and $M_2$ ($M_2\sim 6.3\cdot M_1$ at the electroweak scale).
In this case, the invisible decay of the $Z$ boson, $Z\to\chi_1^0\chi_1^0$, and the chargino production measurements exclude a corner in the parameter space corresponding to low values for $M_1$ and $\mu$ \cite{Bernal:2007uv}.\\
In the scenario {\bf 75} one has $M_1:M_2:M_3\,\sim\,1:-1.2:-1.5$ at the weak scale, so that the LSP is in general close in mass to the lightest chargino and the next-to-lightest neutralino, and hence coannihilation of these states plays a very important role.
There is a thin line up to $\mu$, $M_1\lesssim 1$ TeV where the LSP is dominantly bino-like and the WMAP $\Omega_{DM}\,h^2$ range is obtained by an efficient annihilation and coannihilation of $\chi_1^0$, $\chi_2^0$ and $\chi_1^\pm$.
Moreover, another region giving rise to the correct relic density corresponds to a large band with $\mu\sim 1$ TeV and $M_1\gtrsim 1$ TeV, where the LSP is almost a pure higgsino state.
Once again, coannihilation leads the dark matter relic density.\\
In scenario {\bf 200} the regions fulfilling the dark mater relic density compatible with WMAP constraints correspond to a LSP with mass $\gtrsim 1$ TeV \cite{Bernal:2007uv}.
Owing to the fact that such a heavy candidate is largely outside capabilities of the next generation dark matter direct detection experiments, such scenario will not be considered in the analysis presented in sections \ref{ddp} and \ref{rp}.

\section{Direct detection}\label{dd}
\subsection{Differential event rate}
\label{Xenon}

In spite of the experimental challenges, a number of efforts worldwide
are actively pursuing to directly detect WIMPs with a variety of targets
and approaches. Many direct dark matter detection experiments are
now either operating or in preparation.
All these experiments measure the number $N$ of elastic
collisions between WIMPs and target nuclei in a detector,
per unit detector mass and per unit of time, as a function of the
nuclear recoil energy $E_r$.
The detection rate in a detector depends on the density
$\rho_0\simeq0.3$ GeV cm$^{-3}$ and velocity distribution $f(v_\chi)$ of WIMPs near the Earth.
Usually, the motion of the Earth is neglected and a Maxwellian halo for WIMP's velocity is assumed.
In general, the differential event rate per unit detector mass and
per unit of time can be written as:
\begin{equation}
\frac{dN}{dE_r}=\frac{\sigma_{\chi-N}\,\rho_0}{2\,m_r^2\,m_\chi}\,
F(E_r)^2\int_{v_{min}(E_r)}^{\infty}\frac{f(v_\chi)}{v_\chi}dv_\chi\,,
\label{Recoil}
\end{equation}
where the WIMP-nucleus scattering cross-section, $\sigma_{\chi-N}$, is related to the
WIMP-nucleon cross-section, $\sigma_{\chi-p}$, by $\sigma_{\chi-N}=\sigma_{\chi-p}\cdot (A\,m_r/M_r)^2$, with $M_r=\frac{m_\chi\,m_p}{m_\chi+m_p}$ the WIMP-nucleon reduced mass, $m_r=\frac{m_\chi\,m_N}{m_\chi+m_N}$ the
WIMP-nucleus reduced mass, $m_\chi$ the WIMP mass, $m_N$ the nucleus mass,
and $A$ the atomic weight. $F$ is the nuclear form factor; in the following analysis the Woods-Saxon Form factor will be used.
The integration over velocities is limited to those which
can give place to a recoil energy $E_r$, thus there is a minimal velocity
given by $v_{min}(E_r)=\sqrt{\frac{m_N\,E_r}{2\,m_r^2}}$.\s

In order to compare the theoretical signal with the background 
it is necessary to calculate the $\chi^2$. Let us call $N^{sign}$ 
the signal, $N^{bkg}$ the background and $N^{tot}=N^{sign}+N^{bkg}$ 
the total signal measured by the detector. We will divide the energy 
range between $4$ and $30$ keV in $n=7$ equidistant energy bins. 
For the discrimination between the signal and the background we 
calculate the variance $\chi^2$:
\begin{equation}
\chi^2=
\sum_{i=1}^n\left(\frac{N_i^{tot}-N_i^{bkg}}{\sigma_i}\right)^2\ .
\end{equation}
Here we are assuming a Gaussian error $\sigma_i=\sqrt{\frac{N_i^{tot}}{M\cdot T}}$ 
on the measurement, where $M$ is the detector mass and $T$ the exposure time.
We use a similar analysis to that performed in references \cite{Bernal:2008zk,Bernal:2008cu}.


\subsection{A Xenon-like experiment}

The Xenon experiment aims 
at the direct detection of dark matter via its elastic scattering off 
xenon nuclei. 
It allows the simultaneous measurement
of direct scintillation in the liquid and of ionization,
via proportional scintillation in the gas. In this way, Xenon discriminates
signal from background for a nuclear recoil energy as small as $4.5$ keV.
Currently a $10$ kg detector is being used, but the final 
mass will be $1$ ton of liquid xenon.
\begin{figure}[!ht]
\centering
\includegraphics[width=0.4\textwidth,clip=true,angle=-90]{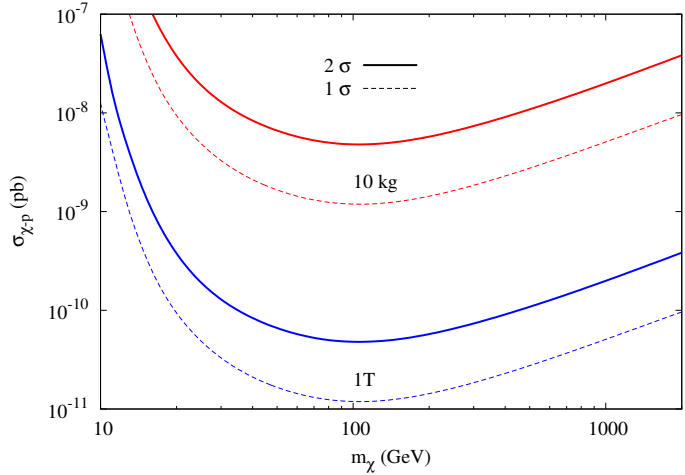}
\vspace{-0.4cm}
\caption{Spin-independent WIMP-nucleon cross-section versus WIMP mass for $\chi^2=1$, $4$  and $M=10$ kg and $1$ ton.}
\label{fig:Xenonsensitivity}
\end{figure}
In figure \ref{fig:Xenonsensitivity}, we show the sensitivity curve for \textit{Xenon10}
 ($M=10$ kg) and \textit{Xenon1T} ($M=1$ ton) for $T=3$ years of data acquisition.\s

In our study, following reference \cite{Angle:2007uj}
we will always consider $7$ energy bins between $4$ and $30$ keV and $3$ years of data acquisition for
a $100$ kg Xenon experiment.
We could take into
account non-zero background using simulations of the recoil spectra of
neutrons in our analysis, and this would significantly degrade the sensitivity of the detector. However, this would involve a much more detailed
study of the detector components (shielding, etc.), and we will not carry it out. In that sense,
our results will be the most optimistic ones.
Comprehensive studies about the influence of astrophysical and background assumptions can be found in references \cite{Bernal:2008zk,Green:2008rd}.\s

\begin{figure}[!ht]
\centering
\includegraphics[width=0.5\textwidth,clip=true,angle=-90]{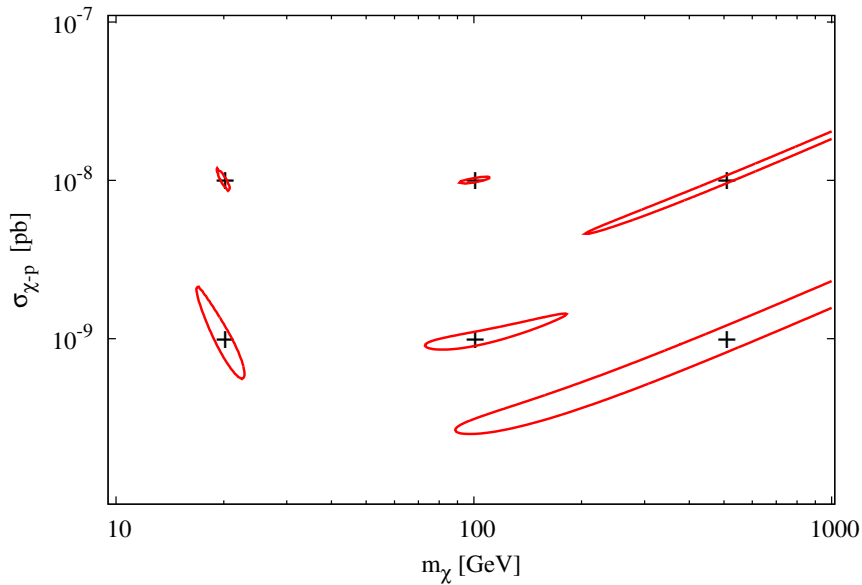}
\vspace{-0.6cm}
\caption{Distribution of the maximum $\chi^2$ in the $[\sigma_{\chi-p},\,m_\chi]$ plane, for $3$ years of exposure in a $100$ kg Xenon experiment, for $m_\chi=20$, $100$, $500$ GeV and $\sigma_{\chi-p}=10^{-8}$, $10^{-9}$ pb.
The lines represent the $68\%$ CL region, and 
the crosses denote the theoretical input parameters.}
\label{patates}
\end{figure}

Figure \ref{patates} shows the ability of Xenon to determine the mass and scattering cross-section for a $20$, $100$ and $500$ GeV WIMP with cross-sections of $10^{-8}$ and $10^{-9}$ pb, in a microscopically model independent approach.
We can clearly see how sensitive the experiment is to light WIMPs: the precision can reach the percent level for $m_\chi\lesssim 50$ GeV and high $\sigma_{\chi-p}$.
For WIMPs much heavier than the nucleus mass ($\sim 100$ GeV for Xenon), the differential rate in equation \eqref{Recoil} becomes almost independent on $m_\chi$ and therefore the relative errors enlarge.
However, for smaller values of the scattering cross-section $\lesssim 10^{-9}$, the possibility of reconstructing $m_\chi$ or $\sigma_{\chi-p}$, at least in a model independent framework, quickly vanishes because of the dramatic increase of the errors.
One way to deal with this deterrent is e.g. to place ourselves into the framework of a given model so as to limit the phase space (i.e. the $(m_\chi\,,\sigma_{\chi-p})$ pairs) upon the inclusion of the available theoretical and experimental constraints.
In this vein, hereafter we leave aside the model independent approach to the dark matter direct detection in order to focus to the MSSM with heavy scalars framework.

\section{Direct detection prospects within the MSSM with heavy scalars}\label{ddp}
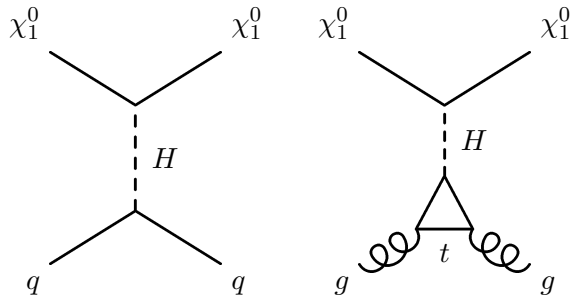
\begin{figure}[!ht]
\vspace{0.5cm}
\centering
\begin{fmfgraph*}(80,80)
  \fmfleftn{i}{2}
  \fmfrightn{o}{2}
  \fmf{plain}{i2,v2,o2}
  \fmf{dashes,label=$H$}{v1,v2}
  \fmf{plain}{i1,v1,o1}
  \fmflabel{$\chi_1^0$}{i2}
  \fmflabel{$\chi_1^0$}{o2}
  \fmflabel{$q$}{i1}
  \fmflabel{$q$}{o1}
\end{fmfgraph*}
\hspace{1cm}
\begin{fmfgraph*}(80,80)
  \fmfleftn{i}{2}
  \fmfrightn{o}{2}
  \fmf{plain}{i2,v2,o2}
  \fmf{dashes,label=$H$,tension=1.5}{v1,v2}
  \fmf{plain}{v3,v1,v4}
  \fmf{plain,label=$t$}{v3,v4}
  \fmf{curly,tension=1.5}{i1,v4}
  \fmf{curly,tension=1.5}{v3,o1}
  \fmflabel{$\chi_1^0$}{i2}
  \fmflabel{$\chi_1^0$}{o2}
  \fmflabel{$g$}{i1}
  \fmflabel{$g$}{o1}
\end{fmfgraph*}
\vspace{.4cm}
\caption{Feynman diagrams contributing to the spin-independent elastic scattering of neutralinos with quarks and gluons, within the MSSM with heavy scalars.}
\label{diffHS}
\end{figure}

As far as the DM direct detection concerns,
the cross-section for elastic scattering of a WIMP with a nuclei detector is perhaps one of the most important properties.
This cross-section determines the detection rate in direct-detection experiments.
The WIMP-nucleus elastic scattering cross-section depends fundamentally on the WIMP-quark interaction strength.\\
In the MSSM with heavy scalars, the leading processes which give rise to the neutralino-nucleus spin-independent interaction are shown in figure \ref{diffHS}.
The diagrams correspond to the elastic scattering of a LSP from a quark and a gluon by the exchange of a Higgs boson in the $t$- and $u$-channels.
The Higgs boson-gluon interaction is induced at the quantum level primarily by top-quark loops.
Since both of the diagrams depend on the interaction between the lightest neutralino and the Higgs boson, the $\chi\chi H$ vertex will drive the phenomenology of the scattering cross-section.
This coupling is given by:
\begin{equation}
C_{\chi_1^0\chi_1^0H}\,\,\propto\,\, N_{13}\,\left(\gtild\,N_{12}-\gtildp\,N_{11}\right)-N_{14}\,\left(\gtilu\,N_{12}-\gtilup\,N_{11}\right)\,.
\label{coupling}
\end{equation}
As equation \eqref{coupling} depends on the medley of the whole $N_{ij}$ matrix elements, 
it follows that the $H\chi\chi$ coupling can be enhanced for a lightest neutralino which is a `temperate' gaugino-higgsino mixture.
In addition, for a pure higgsino-like or a pure gaugino-like neutralino, the coupling with a Higgs boson vanishes, and hence it does the scattering cross-section.\\
In the general MSSM case, there also exists a potentially important input to the cross-section coming from a plethora of diagrams involving squarks propagators (essentially the first generation squarks $\tilde u$ and $\tilde d$), both contributing to the interaction of WIMPs with quarks and with gluons \cite{Jungman:1995df}.
These processes could be dominant in the case when the squarks are light.
Nevertheless, in this case we are taking into account heavy scalars, hence diagrams containing squarks will be suppressed by powers of $1/\msusy^2$.\\
In the same vein, let us recall that the $Z$-boson exchange in the interaction $\chi_1^0 q\to\chi_1^0 q$ does not contribute to the spin-independent scattering cross-section but only to the spin–dependent one.
As already pointed out \cite{Profumo:2004at}, spin–dependent direct neutralino searches in the general MSSM appear to be largely disfavored with respect to spin-independent ones.\s

\begin{figure}[!ht]
\centering
\includegraphics[width=0.5\textwidth,clip=true,angle=-90]{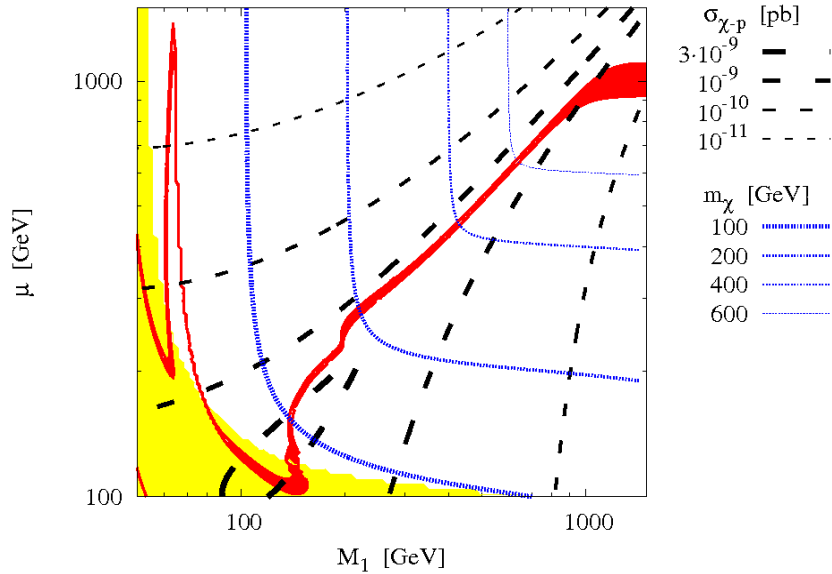}
\vspace{-0.6cm}
\caption{Contour lines for the scattering cross-section $\sigma_{\chi-p}$ (black dashed lines) and the mass of the LSP (blue dotted lines) in the $[M_1,\,\mu]$ plane, for $\msusy=10^4$ GeV, $\tb=5$ and the universal scenario.
The regions fulfilling the WMAP constraint (red--dark gray) and excluded by collider searches (green--light gray) are also shown.}
\label{iso30-4}
\end{figure}

In the same way as for the dark matter relic density, the computation of scattering cross-sections has been performed using an adapted version of the micrOMEGAs code.
Thus, it is possible to use equation \eqref{Recoil} in order to extract the expected differential event rate of nuclear recoil, in the framework of the Xenon experiment.
Figure \ref{iso30-4} presents the contour lines for the scattering cross-section $\sigma_{\chi-p}$ (dashed lines) and the mass of the lightest neutralino (dotted lines) in the $[M_1,\,\mu]$ plane, for $\msusy=10^4$ GeV, $\tb=5$ and the universal scenario.
It is also shown the region allowed by the WMAP measurements and the parameter space excluded by collider data.
Worth noticing is the fact that while $M_1\sim\mu$, the cross-section reaches high values, up to $\sim 4\cdot 10^{-9}$ pb.
The $\chi\chi H$ coupling stays high even for elevated $M_1$ and $\mu$ values, but the maximum slowly narrows.
However, as we move away from the higgsino-gaugino mixing region, $\sigma_{\chi-p}$ decreases severely.
Let us emphasize that in the Higgs-pole region, the bino-like nature of the lightest neutralino implies a highly suppressed scalar cross-section via Higgs boson exchange because of the low $\chi\chi H$ coupling.
In that sense, the Higgs boson resonance does not enhance the scattering cross-section $\sigma_{\chi-p}$.\\
Changing $\msusy$ to higher values causes a rapid decrease in the scattering cross-section.
This may be largely explained by the variation of the Higgs boson mass, which increases logarithmically with $\msusy$, and by the strong dependence of the cross-section $\sigma_{\chi-p}$ on the mass $M_H$:
\begin{equation}
\sigma_{\chi-p}\propto\frac{1}{M_H^4}\,.
\end{equation}
On the other hand, the increase of $\tan\beta$ reduces the scattering cross-section because of two effects: i) the slight growth of the Higgs mass and ii) the LSP becomes more quickly a pure bino- or higgsino-like state.
Smaller values for $\tb$ ($\lesssim 2-3$) produce a SM-like Higgs boson which becomes too light, actually excluded by negative searches at LEP2.
In this vein, the analysis in the next section will be focused to the kind of best scenario for dark matter direct detection, i.e. $\msusy=10^4$ GeV and $\tb=5$.\s

\begin{figure}[ht!]
\centering
\hspace{-0.7cm}\includegraphics[width=7.0cm,clip=true,angle=-90]{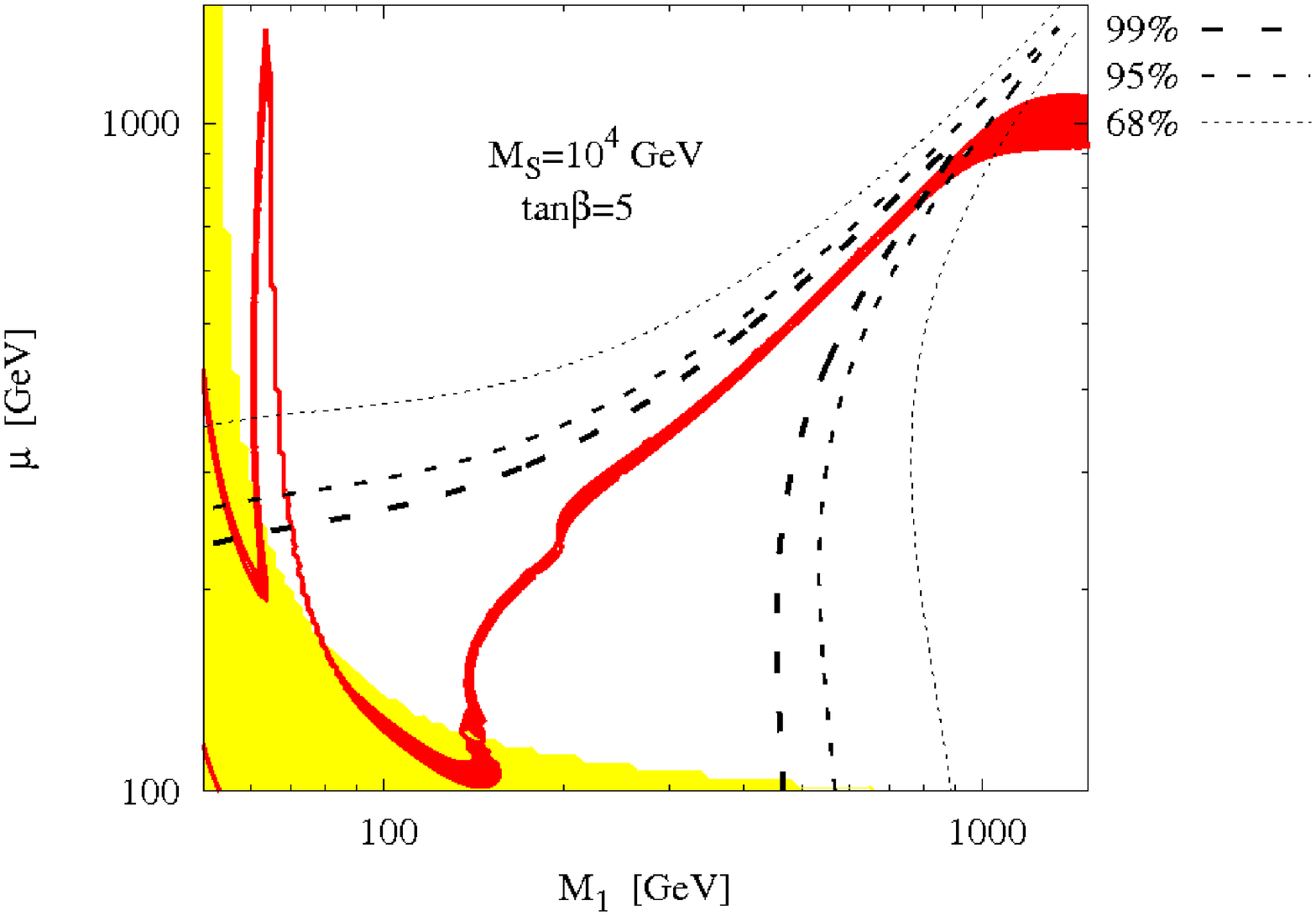}\\\vspace{-0.4cm}
\hspace{-0.7cm}\includegraphics[width=7.0cm,clip=true,angle=-90]{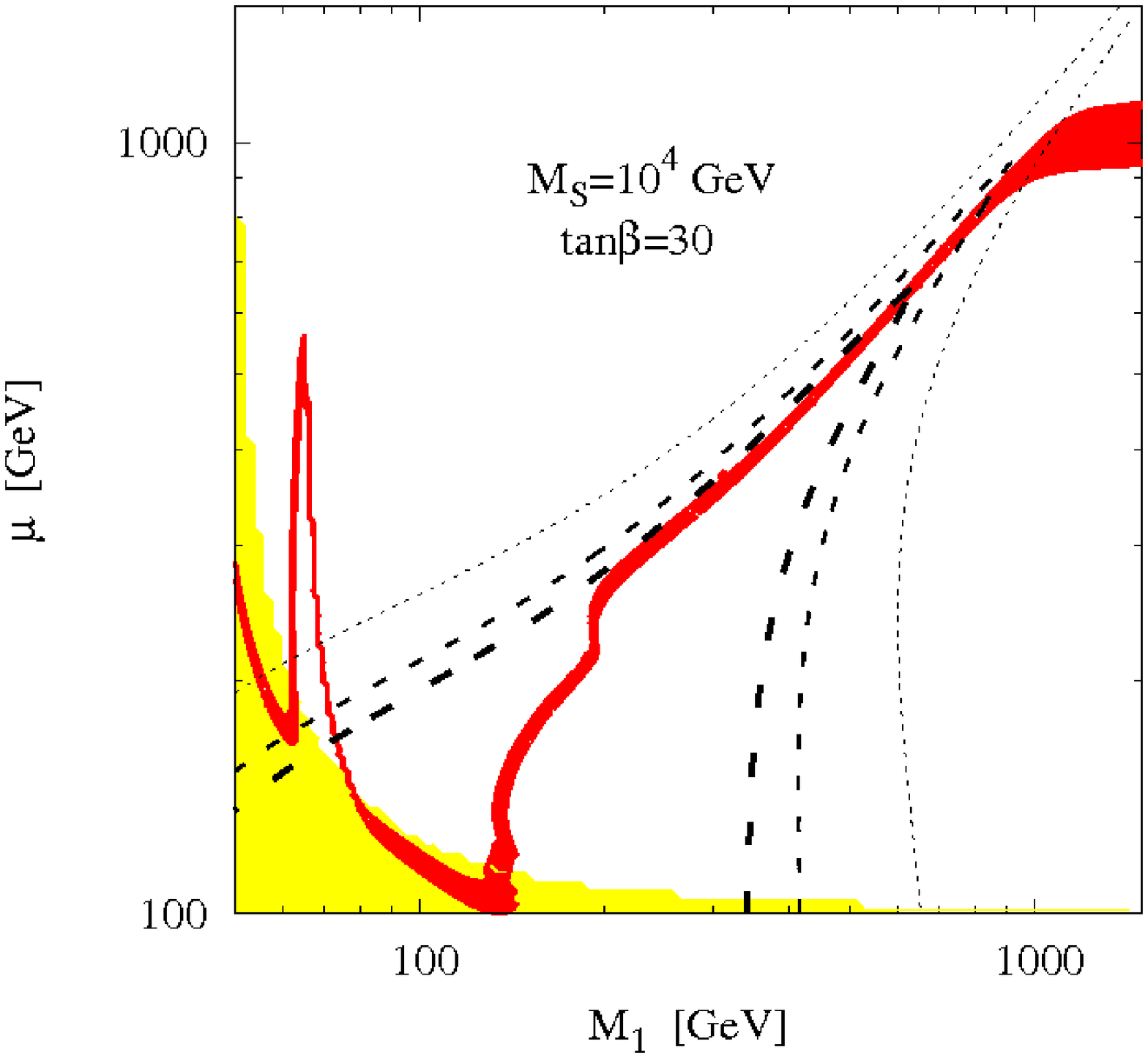}
\hspace{ 1.0cm}\includegraphics[width=7.0cm,clip=true,angle=-90]{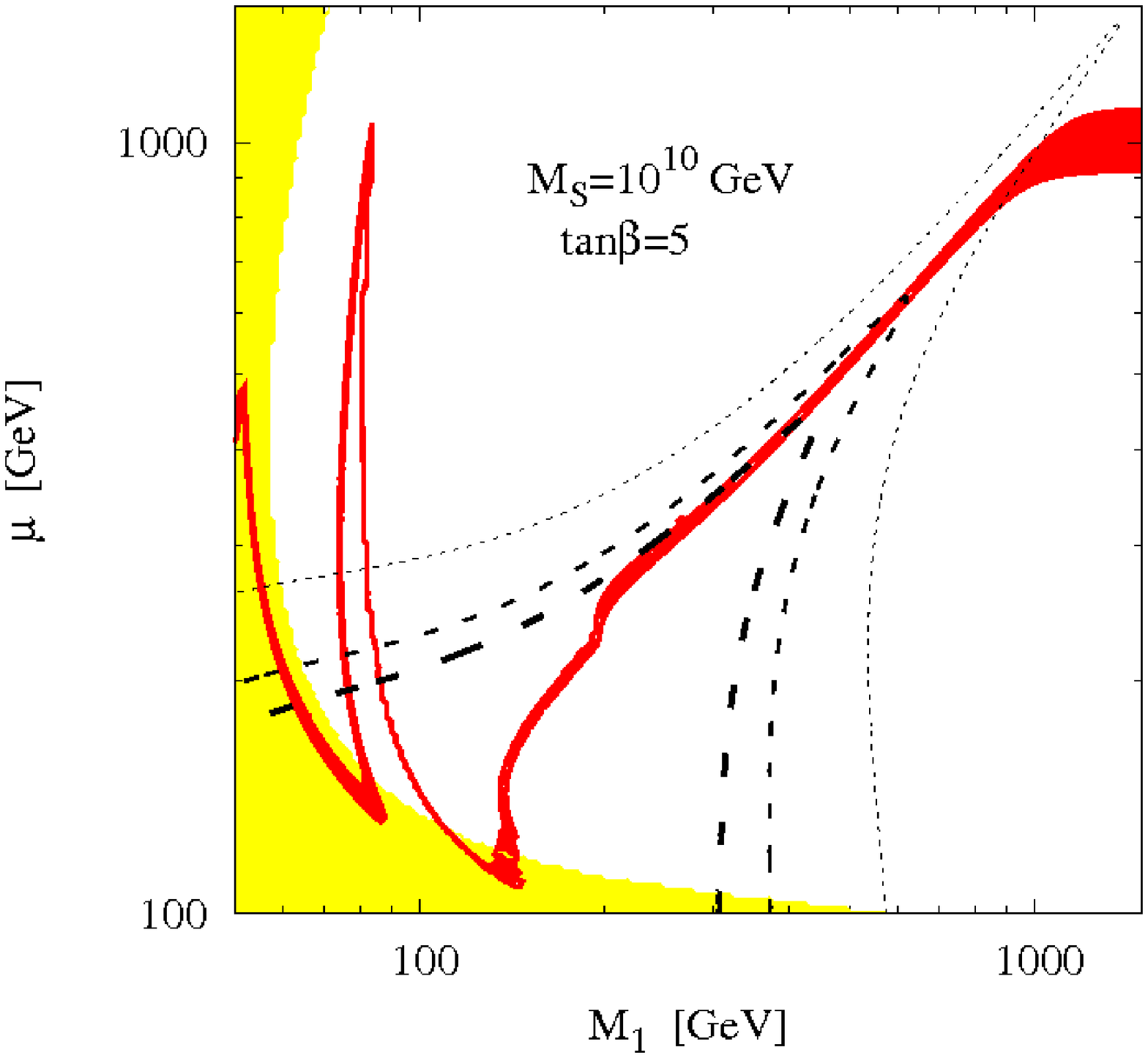}
\vspace{-0.6cm}
\caption{Exclusion lines for $3$ years of exposure in a $100$ kg Xenon experiment, on the $[M_1,\,\mu]$ plane, for different combinations of $\msusy$ and $\tb$, in the framework of the universal scenario.
The dashed lines represent the $68\%$, $95\%$ and $99\%$ CL exclusion lines.
The regions fulfilling the WMAP constraint (red--dark gray) and excluded by collider searches (green--light gray) are also shown.}
\label{exclu30-4}
\end{figure}

The maximum scattering cross-section reached in figure \ref{iso30-4} is not yet ruled out by dark matter direct detection experiments;
however, some collaborations aim to exclude cross-sections up to $10^{-10}$ pb in a not too distant future.
In figure \ref{exclu30-4} we show exclusion lines for Xenon after $3$ years of data acquisition, on the $[M_1,\,\mu]$ parameter space, in the framework of the universal scenario.
These curves reflect the Xenon sensitivity and represent its ability to test and exclude different regions of the MSSM with heavy scalars at $68\%$, $95\%$ and $99\%$ CL.
The regions which could be excluded follow approximately the contour lines for the scattering cross-section (see figure \ref{iso30-4}).
Moreover, they tend to enlarge for LSP masses of the order of $m_\chi\sim 100$ GeV, as expected from figure \ref{fig:Xenonsensitivity}.
As can be seen in figure \ref{exclu30-4}, for the universal scenario the absence of signal at Xenon could exclude a sizeable fraction of the parameter space of the MSSM with heavy scalars, in particular a major part of the region fulfilling the WMAP constraint.
In fact, only the top of the Higgs peak and the area corresponding to $\mu\sim 1$ TeV and $M_1\gtrsim 1$ TeV could not be probed.
This is due to the low scattering cross-section and the high LSP mass respectively.\s

\begin{figure}[ht!]
\centering
\hspace{-0.7cm}\includegraphics[width=7.0cm,clip=true,angle=-90]{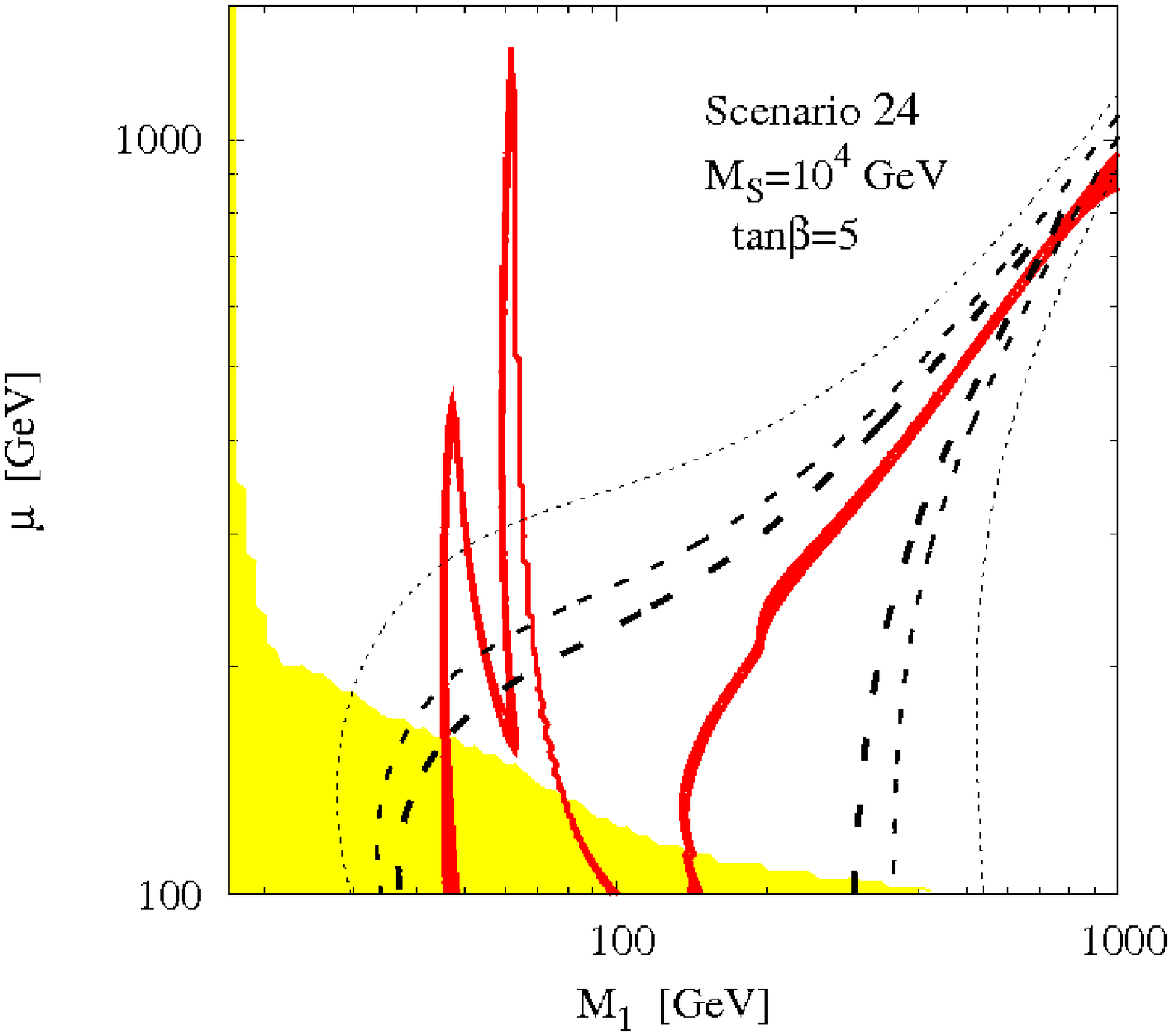}
\hspace{ 1.0cm}\includegraphics[width=7.0cm,clip=true,angle=-90]{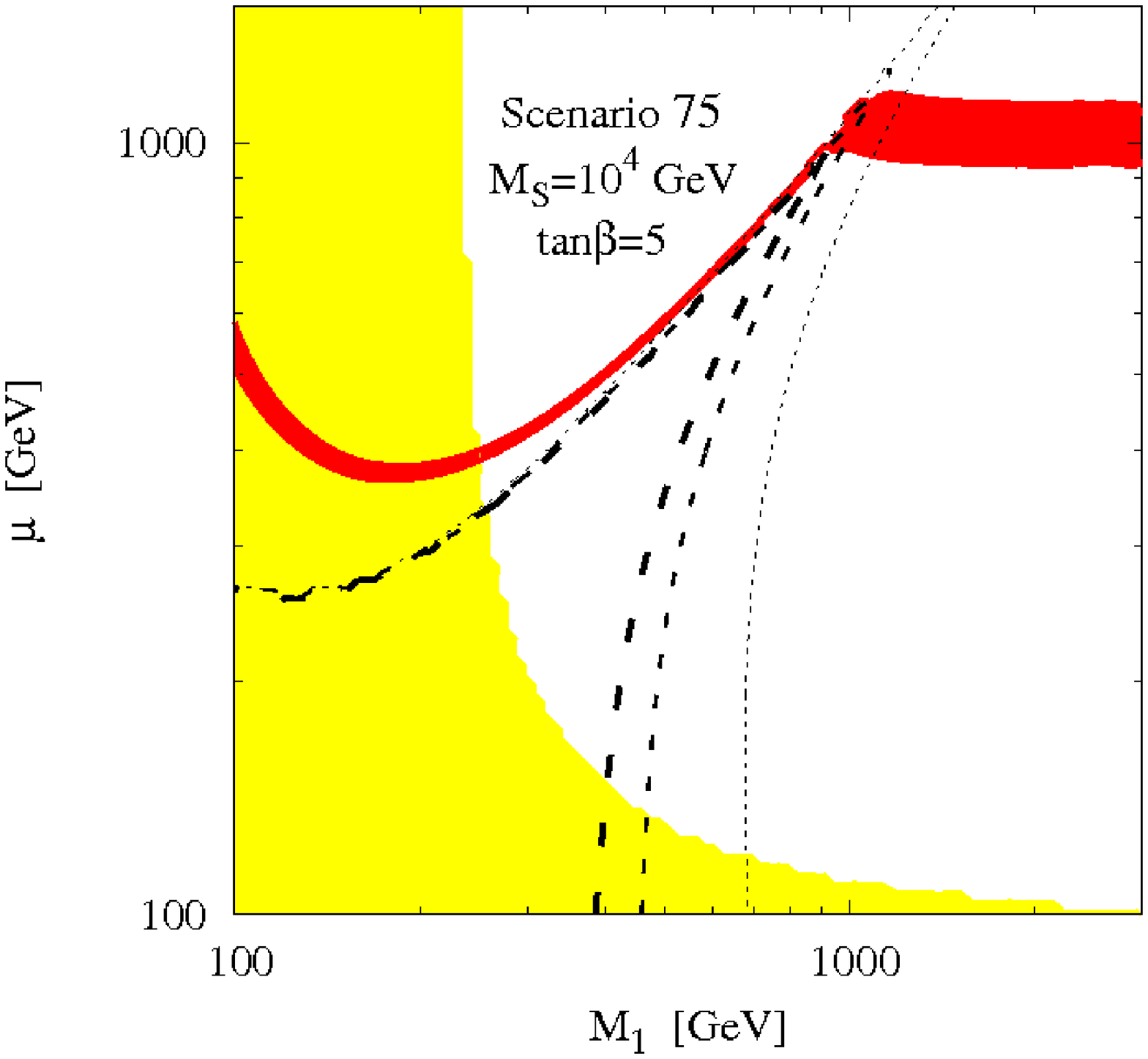}
\vspace{-0.6cm}
\caption{The same as figure \ref{exclu30-4} but for scenarios {\bf 24} (left pane) and {\bf 75} (right pane).}
\label{exclunon}
\end{figure}

Figure \ref{exclunon} is similar to the previous one, with the difference that we explore the non-universal scenarios {\bf 24} and {\bf 75}.
In the same way as for the universal case, for scenario {\bf 24} the exclusion lines follow approximately the $\mu\sim M_1$ line, getting large for WIMP masses of the order of $100$ GeV.
For $m_\chi\lesssim 80$ GeV, the sensitivity of the experiment deteriorates rapidly (see figure \ref{fig:Xenonsensitivity}).
Let us note that the dark matter scattering cross-section is not enhanced by the resonant exchange of a $Z$ boson, in the same way as for the Higgs boson peak.
Finally, the direct detection prospects for scenario {\bf 75} are not encouraging.
In fact, in this case the regions leading to the correct dark matter relic density measured by WMAP are not generated by the usual annihilation process, but rather by coannihilation processes of the LSP with the $\chi_2^0$ and the $\chi_1^\pm$.
Hence, most of the parameter space region that a Xenon-like experiment could examine corresponds to an annihilation cross-section which turns out to be too high indeed to be measurable. 
Nevertheless, a small single region exists with $\mu\sim M_1\sim 1$ TeV which gives rise to the correct relic density and which could be potentially excluded by a Xenon-like experiment; but such region corresponds to a very heavy LSP, at the actual border of detectability.

\section{Reconstruction prospects}\label{rp}
Once a signal has been detected in a given direct dark matter detection experiment, a new question appears concerning the possibilities of reconstructing both physical observables and fundamental parameters of the model.
Gradually, a new generation of dark matter experiments start to make measurements and not simply to set limits \cite{Hooper:2006wv}.
In this vein, the aim of this section is to assess, first of all, the possibilities of identifying neutralino LSP properties, as its mass and its scattering cross-section, in the framework of the MSSM with heavy scalars from a Xenon-like experiment.
Secondly, we will also examine the prospects of reconstructing the fundamental parameter space, i.e. the low-energy Lagrangian parameters.
Even if in the general MSSM it is a formidable task, the reduced number of free parameters of the present model makes the work more feasible.
However, for reconstructing all parameters using direct detection experiments there is a first difficulty because they only give rise to a single observable: the differential event rate of nuclear recoil.
This can be partially eased by taking into account the existing constraints coming from collider physics but especially the one given by WMAP.
A second complication comes from the fact that the sector of the parameter space that can be tested by Xenon is almost insensitive to some of the parameters.
Actually, $\msusy$ and $\tb$ will essentially determine the mass 
of the SM-Higgs boson, but their impact in the region where the LSP is a higgsino-gaugino mixture is somehow limited (e.g. see figures \ref{dm4}-\ref{dmmas} and \ref{exclu30-4}).
Therefore, we will concentrate on the reconstruction over the $[M_1,\,\mu]$ plane in the framework of a given pattern of soft SUSY-breaking gaugino masses.
Other measurements (e.g. at LHC or ILC) could bring complementary information about the remaining low-energy Lagrangian parameters.\s

In order to study the abilities of Xenon to reconstruct the parameter space of the MSSM with heavy scalars, we have chosen four characteristic benchmark points defined in table \ref{tabben}, corresponding to different values for the $(M_1,\,\mu)$ pair and different scenarios for gaugino masses.
\begin{table}[ht!]
\centering
\renewcommand{\arraystretch}{1.3}
\begin{tabular}{|c||c|cc|cc|}
\hline
   & Scenario & $\mathbf{M_1}$ [GeV] & $\mathbf{\mu}$ [GeV] & $\mathbf{m_\chi}$ [GeV] & $\mathbf{\sigma_{\chi-p}}$ [pb]\\\hline\hline
\textbf{A}  & {\bf 1}  & $138$  & $143$ & $93.6$  & $3.2\cdot 10^{-9}$  \\
\textbf{B}  & {\bf 1}  & $207$  & $262$ & $185.1$ & $1.7\cdot 10^{-9}$  \\
\textbf{C}  & {\bf 1}  & $71$   & $243$ & $63.2$  & $3.0\cdot 10^{-10}$ \\\hline
\textbf{D}  & {\bf 24} & $45$   & $165$ & $39.0$  & $2.9\cdot 10^{-10}$ \\\hline
\end{tabular}
\caption{Benchmark points used throughout the analysis, for $\msusy=10^4$ GeV, $\tb=5$ and different scenarios of gaugino masses.
The corresponding LSP mass and scattering cross-section are also presented.}
\label{tabben}
\end{table}
We have stuck to low values for the parameters $\msusy$ and $\tb$ ($\msusy=10^4$ GeV and $\tb=5$) in order to maximize the direct detection potential of Xenon.
Let us recall that in this limit the scattering cross-section is maximized.
All these points satisfy the whole collider constraints and also generate a dark matter relic density in agreement with WMAP measurements.
Moreover, they belong to the region that could be probed by a $100$ kg Xenon experiment after $3$ years of exposure.
For each point, table \ref{tabben} also presents the corresponding LSP mass and scattering cross-section.\\
Points {\bf A} and {\bf B} correspond to the mixed region in which the lightest neutralino is a higgsino-bino mixture and then, they give rise to a high scattering cross-section $\sigma_{\chi-p}\gtrsim 10^{-9}$ pb.
However, from the point of view of direct detection, point {\bf A} presents an important advantage over point {\bf B}.
In fact, point {\bf A} generates a neutralino with mass $m_\chi\sim 100$ GeV, corresponding to the maximal sensitivity, while point {\bf B} produces a heavier LSP.
In the latter case, the dark matter relic density is mainly produced by the reaction $\chi_1^0\,\chi_1^0\to t\,\bar t$.
On the other hand, point {\bf C} corresponds to the Higgs pole region in which the lightest neutralino is rather light, $m_\chi\sim \frac12 M_H\sim 65$ GeV.
Despite its low mass, this point gives rise to a very weak scattering cross-section $\sigma_{\chi-p}\sim 10^{-10}$ pb, in the threshold of detectability.
Finally, point {\bf D} is located over the $Z$-pole peak and corresponds to an LSP with a very light mass ($m_\chi\sim \frac12 M_Z\sim 45$ GeV) and scattering cross-section.
Because of the large mass splitting between the bino and the wino mass terms in scenario {\bf 24}, $M_2\sim 6.3\cdot M_1$ at the electroweak scale, such an LSP mass is not excluded.\s

\begin{figure}[ht!]
\centering
\hspace{-1.7cm}\includegraphics[width=7.0cm,clip=true,angle=-90]{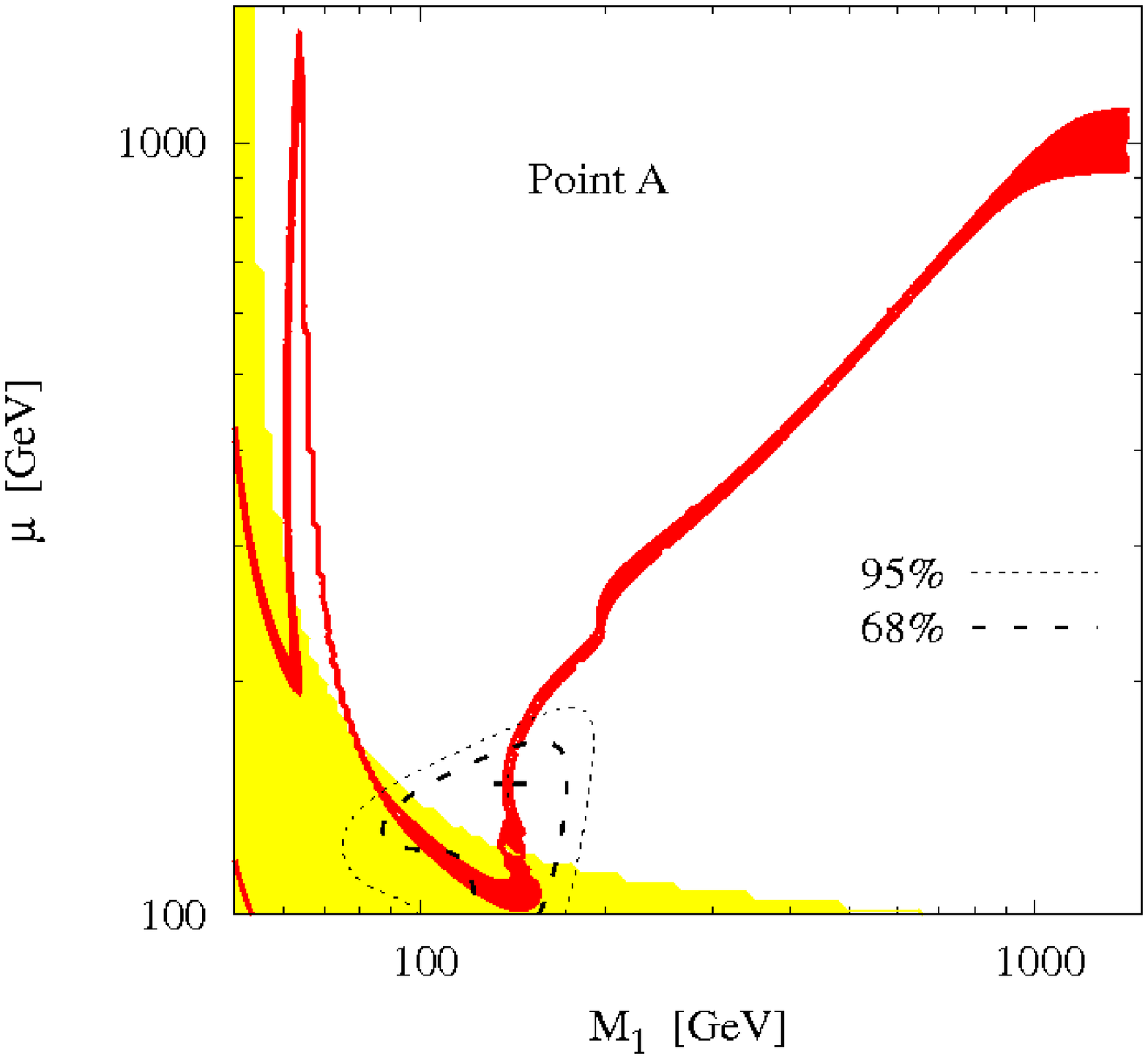}
\hspace{ 1.0cm}\includegraphics[width=7.0cm,clip=true,angle=-90]{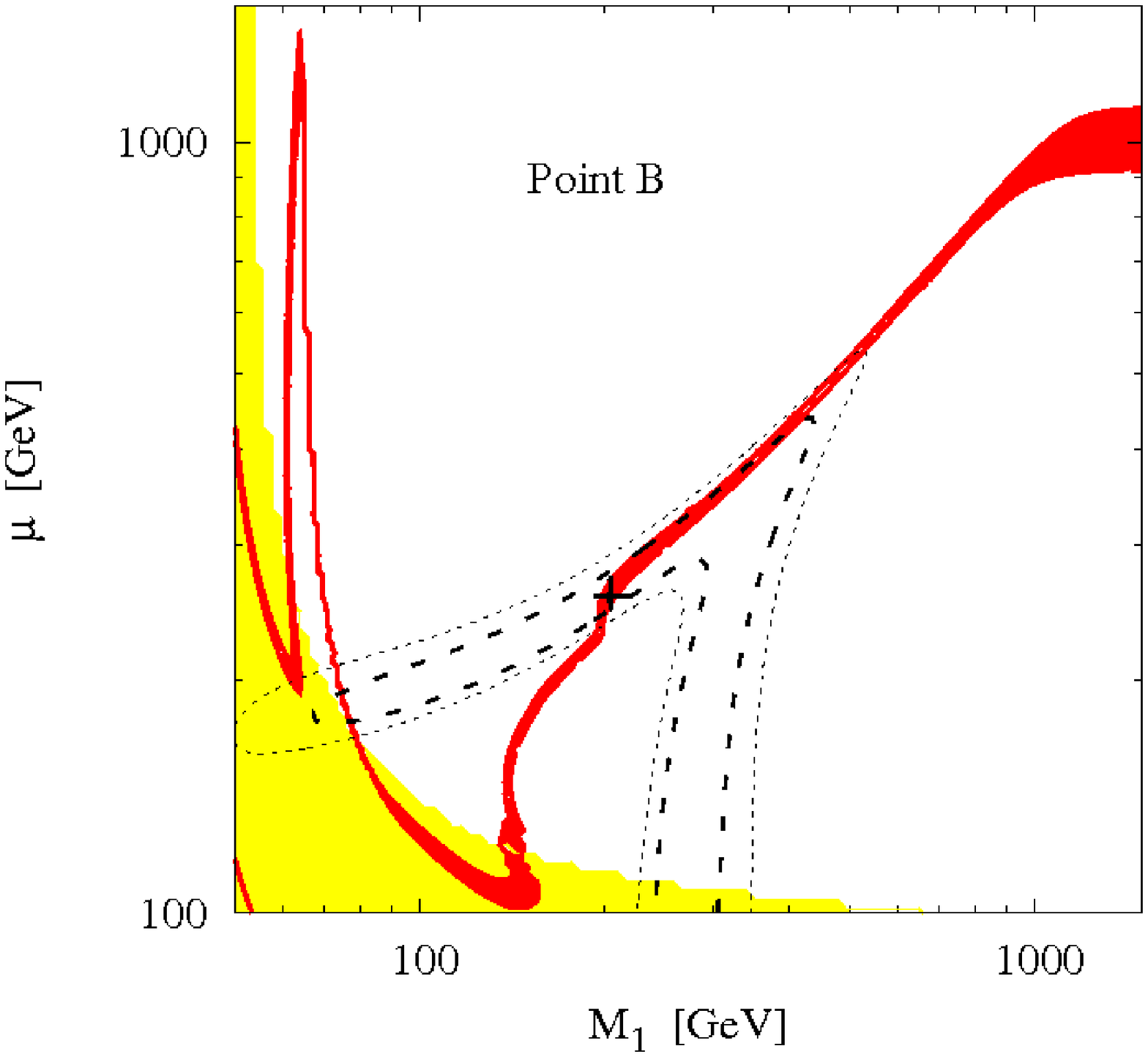}\\\vspace{-0.42cm}
\hspace{-1.5cm}\includegraphics[width=7.0cm,clip=true,angle=-90]{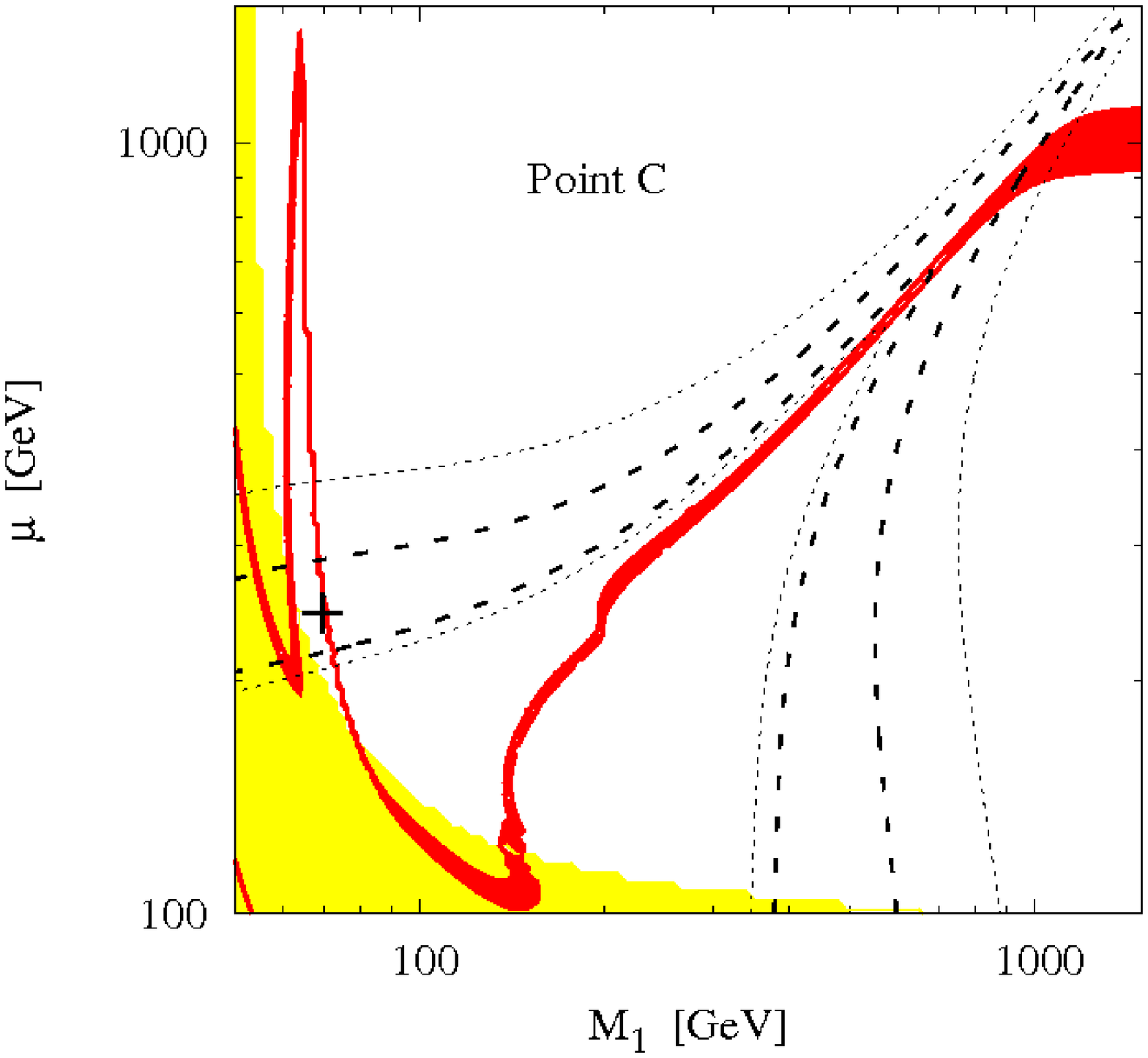}
\hspace{ 1.1cm}\includegraphics[width=7.0cm,clip=true,angle=-90]{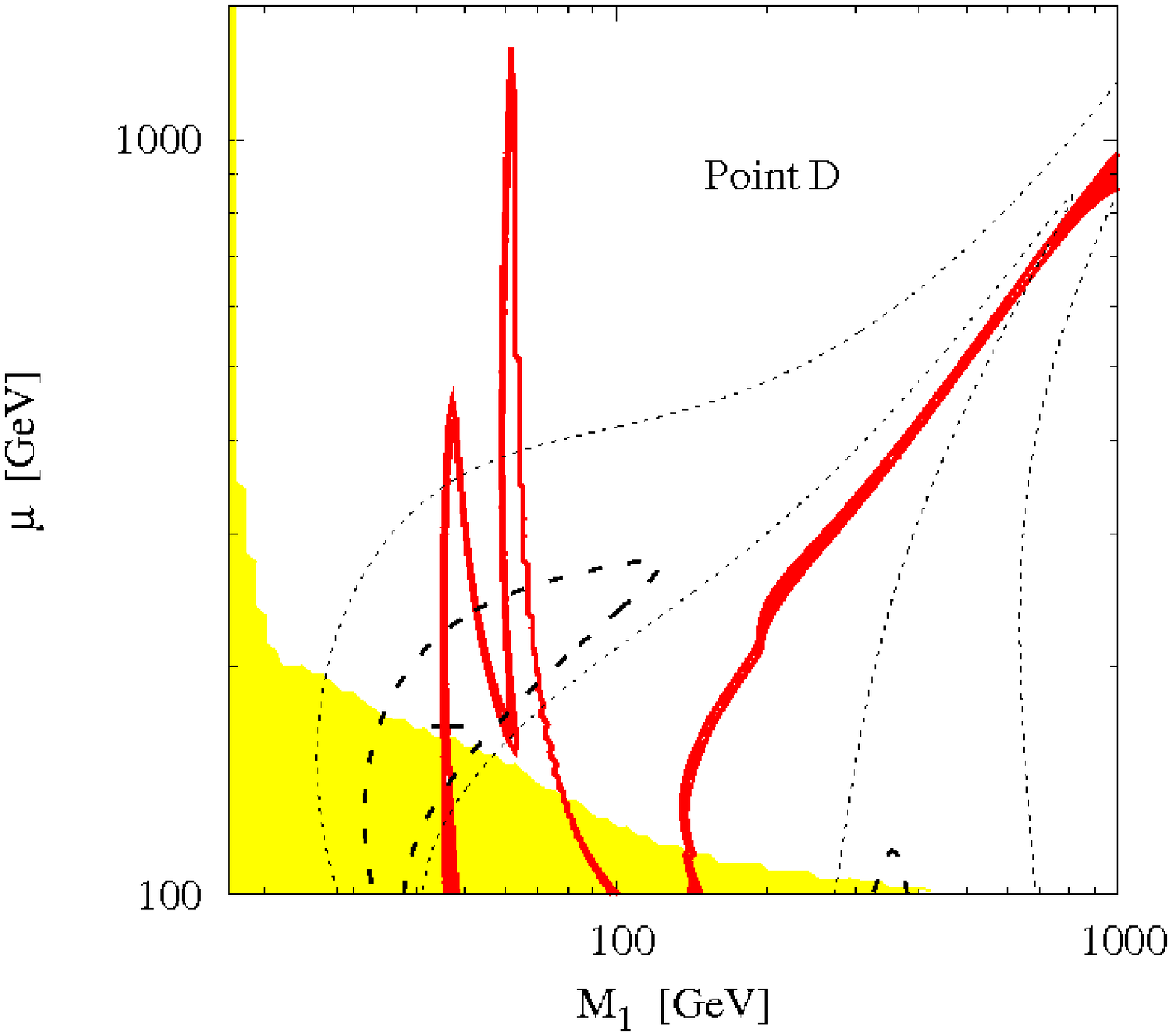}
\vspace{-0.42cm}
\caption{Ability of a $100$ kg Xenon experiment after $3$ years of exposure to reconstruct the $M_1$ and $\mu$ parameters, for the three benchmarks of the table \ref{tabben}, at $68\%$ and $95\%$ CL.
The crosses denote the theoretical input parameters.
The regions fulfilling the WMAP constraint (red--dark gray) and excluded by collider searches (green--light gray) are also shown.}
\label{pata30-4}
\end{figure}

Figure \ref{pata30-4} shows the ability of Xenon to reconstruct the $M_1$ and $\mu$ parameters, for the four benchmarks described in table \ref{tabben}.
Even if the reconstructed regions are in general relatively large, in particular for points {\bf B} and {\bf C}, the conjunction with collider and cosmological constraints allows to drastically shrink the latter.\\
For point {\bf A}, for instance, the overlap of the different bounds strongly restricts the region that cannot be discriminated by Xenon in the [$M_1,\,\mu$] plane to $M_1\sim 141$ GeV and $121$ GeV $\lesssim\mu\lesssim 165$ GeV.
Henceforth, the regions will correspond to a $68\%$ CL.
This range corresponds to a lightest neutralino with a mass within the interval $84$ GeV $\lesssim m_\chi\lesssim 104$ GeV and a cross-section $\sigma_{\chi-p}$ between $2.9\cdot 10^{-9}$ pb and $3.4\cdot 10^{-9}$ pb.
Note that, for this case, both the LSP mass and the scattering cross-section can be very well reconstructed with a precision of the order of $20\%$ and $15\%$ respectively.\\
With regard to points {\bf B} and {\bf C}, they present highest values for the LSP mass or the cross-section implying a spoilage of the reconstruction capacities.
Actually, for those points, the discrimination regions grow and branch in two areas almost symmetrical about the $\mu=M_1$ axis.
The reconstructed area compatible with constraints for point {\bf B} consists of two parts.
The largest one almost follows the $\mu=M_1$ line with $M_1$ varying within the interval delimited by $197$ GeV and $377$ GeV corresponding to a LSP  with a well defined scattering cross-section of the order of $\sim 2\cdot 10^{-9}$ pb but with a mass within the interval of $173-350$ GeV.
Moreover, the signal is also compatible with a very narrow region located over the Higgs-peak and corresponding to $M_1\sim 75$ GeV and $172$ GeV $\lesssim\mu\lesssim 192$ GeV.
This region gives rise to a $\sim 61$ GeV LSP with a scattering cross-section between $1.1\cdot 10^{-9}$ pb and $7.5\cdot 10^{-10}$ pb.
A $100$ kg Xenon experiment after $3$ years of exposure would not be able to disentangle these two unrelated areas.\\
The region engendered by point {\bf C} is located around $M_1\sim 71$ GeV and $218$ GeV $\lesssim\mu\lesssim 285$ GeV.
In such an interval the mass of the lightest neutralino is close to $61$ GeV but the cross-section can vary in a large range, from $1.5\cdot 10^{-10}$ pb to $4.6\cdot 10^{-10}$ pb.
There is, nevertheless, an extra region compatible both with the dark matter signal and the whole set of constraints, standing near $M_1\sim\mu\sim 0.7-1$ TeV. 
However, even if this area gives rise to a large scattering cross-section of $\sim 2\cdot 10^{-9}$ pb, because of the large mass of the LSP, we are in a regime where the detector is at the limit of its capabilities.\\
Finally, the reconstructed region fulfilling the constraints for point {\bf D} consists of four parts corresponding to the left and right bands of the $Z$- and $H$-peaks.
They are roughly limited by $160$ GeV $\lesssim\mu\lesssim 255$ GeV and $M_1\sim 45$ GeV, $55$-$60$ GeV and $68$ GeV.
In the four areas, the LSP mass varies in the range between $37$ GeV and $62$ GeV; the scattering cross-section varies between $1.2\cdot 10^{-10}$ pb $\lesssim\sigma_{\chi-p}\lesssim 3.4\cdot 10^{-10}$ pb.
For this case, the LSP mass can be reconstructed with a precision of the order of $65\%$, whereas, for the cross-section the relative error reaches almost $75\%$.
It is clear that a Xenon-like experiment cannot examine with a high-precision level such a scenario with a so low scattering cross section.
However, it can provide very valuables hints on the nature of the WIMP dark matter.



\section{Conclusions}\label{conclu} 
We have explored the dark matter detection prospects in the Minimal Supersymmetric
Standard Model in the scenario where the scalar partners of the
fermions and the Higgs particles (except for the Standard-Model-like
one) are assumed to be very heavy and are removed from the low-energy
spectrum.
In the MSSM with heavy scalars, the WIMP candidate for the dark matter relic density is the neutralino LSP.
We have analysed the neutralino LSP ($\chi_1^0$)
in scenarios where the gaugino mass parameters are universal but also the case where they are non-universal at the GUT scale.
This analysis has been carried out in the framework of a Xenon-like $100$ kg experiment.
In general, an important fraction of the parameter space giving rise to the dark matter relic density measured by WMAP can be probed and excluded in the case of not detecting any WIMP.
In the opposite case, once a WIMP signal has been found, we have shown that for a light $\chi_1^0$ which is a higgsino-gaugino mixture it is possible to reconstruct efficiently the mass and the scattering cross-section of the neutralino LSP.
Moreover, we have shown that it is also feasible to put strong constraints over some of the parameters of the Lagrangian, e.g. the higgsino and the gaugino mass parameters.

\subsection*{Acknowledgments:} 
We would like to thank A. Djouadi, A. Goudelis, D. López-Val and Y. Mambrini for helpful discussions and for careful reading of the manuscript.
NB thanks an ESR position of the EU project RTN MRTN-CT-2006-035505 HEPTools.

\bibliographystyle{utphys}
\addcontentsline{toc}{section}{Bibliography}
\bibliography{bibDMSplit}

\end{fmffile}
\end{document}